\documentclass[]{aa}  

\usepackage{graphicx}
\usepackage{natbib}
\usepackage{txfonts}
\usepackage[utf8]{inputenc}

\begin{document}

\title{Dark Matter Response to Galaxy Assembly History}

\author{M. Celeste Artale\inst{1}\thanks{E-mail:Maria.Artale@uibk.ac.at, mcartale@gmail.com} 
\and Susana E. Pedrosa\inst{2}
\and Patricia B. Tissera\inst{3}
\and Pedro Cataldi\inst{4} 
\and Arianna Di Cintio\inst{5,6} \\
}

\institute{Institut f\"{u}r Astro- und Teilchenphysik, Universit\"{a}t Innsbruck, Technikerstrasse 25/8, 6020 Innsbruck, Austria 
\and CONICET-Universidad de Buenos Aires. Instituto de Astronom\'ia y F\'isica del Espacio (IAFE), CC 67, Suc. 28, 1428 Buenos Aires, Argentina
\and Departamento de Ciencias F\'isicas, Universidad Andres Bello, Av. Republica 220, Santiago, Chile
\and Instituto de F\'isica de Buenos Aires, IFIBA, UBA CONICET, Pabellon 1, Ciudad Universitaria, 1428 Buenos Aires, Argentina
\and Instituto de Astrofisica de Canarias, Calle Via L\'actea, La Laguna, Tenerife, Spain 
\and University of La Laguna. Avda. Astrof\'isico Fco. S\'anchez, La Laguna, Tenerife, Spain 
}

   \date{Received XXXX; accepted XXXX}

   \abstract
   {}
   {It is well known that the presence of baryons affects the dark matter host haloes. 
    Exploring the galaxy assembly history together with the dark matter haloes properties through time can provide a way to measure these effects.}
   {We study the properties of four Milky Way mass dark matter haloes from the Aquarius project during their assembly history, between $z=0-4$.
  In this work, we use the SPH run from Scannapieco et al. (2009) and the dark matter only counterpart as case studies. 
  To asses the robustness of our findings, we compare them with one of the haloes run using a moving-mesh technique and different sub-grid scheme.}
   {Our results show that the cosmic evolution of the dark matter halo profiles
   depends on the assembly history of the baryons. We find that the dark
   matter profiles do not significantly change with time, hence they become stable, 
   when the fraction of baryons accumulated in the central regions reaches 80 percent
   of its present mass within the virial radius.
   Furthermore, the mass accretion history shows that the haloes that assembled earlier are those that contain a 
   larger amount of baryonic mass aforetime, which in turn allows the dark matter halo profiles to reach a stable configuration earlier.
   For the SPH haloes, we find that the specific angular momentum
   of the dark matter particles within the five percent of the virial radius at
   z = 0, remains approximately constant from the time at which 60 percent of the stellar mass is gathered.
   We explore different theoretical and empirical models for the contraction of the haloes through redshift. A model to better describe the contraction of the haloes through redshift evolution must depend on the stellar mass content in the inner regions. }
   {}


\maketitle
 
\section{Introduction}
\label{sec:intro}

Within the current standard cosmological model $\Lambda$-CDM, the dark matter together with the dark energy are the main ingredients that shape the formation and evolution of cosmic structures.  
In this scenario, galaxies assemble and evolve within dark matter haloes formed via hierarchical growth \citep{White1978}.

In the past decades, N-body simulations that account only for the dynamics of the cold dark matter, have been carried out 
to investigate the properties of the haloes. Their results show that 
dark matter haloes follow a universal radial density profile 
\citep[NFW,][]{Navarro1997,Diemand2005}, and 
tend to be triaxial in shape, and prolate in the inner regions \citep{Frenk1988,Jing2002,Allgood2006,Stadel2009}.
However, these simulations do not take into account 
the effects of galaxy formation 
and the presence of baryons on the evolution of the dark matter haloes.
Observational results based on rotational curves from spiral galaxies show that dark matter haloes 
might have a central density core \citep[e.g.,][]{Salucci2000,Oh2008,Donato2009}, contrary to 
the \textit{cuspy} universal profiles obtained from simulations that only consider cold dark matter.

Current cosmological hydrodynamical simulations that account for the dark matter halo assembly
together with galaxy formation have proven to be useful to investigate the impact of baryons on different
properties of the dark matter haloes.
For instance, baryons can transform the shape of dark matter haloes into more spherical or axisymmetric configurations
\citep{Tissera1998,Kazantzidis2004,Tissera2010,Bryan2012,Zemp2012,Bryan2013,Ceverino2015,Zhu2017}, 
change the distribution of dark matter in the inner regions \citep{Governato2012,Maccio2012,DiCintio2014b,DiCintio2014,Zhu2016},
and modify their specific angular momentum \citep{Zavala2016}.

It is well established that within galaxy formation models
there are many physical processes that can modify the properties of dark matter haloes.
The inflow of baryons onto the central galaxy contracts the dark matter mass to the inner regions. 
This effect has been often modelled assuming adiabatic contraction models \citep{Blumenthal1986,Gnedin2004}, based on that the
gravitational potential of the system is spherically symmetric and changes slowly. 
However, this simplified model does not reproduce the results obtained from different cosmological hydrodynamical simulations 
and other empirical models have been considered \citep[e.g.,][]{Abadi2010,Pedrosa2010,Chan2015}.
Additionally, the outflows produced by stellar feedback can prevent gas cooling by removing it from the central regions and in turn, 
expanding the haloes \citep{Read2005,Maccio2012,Governato2012,Brook2015,El-Zant2016}.
Other processes might also modify the dark matter haloes such as the presence of galactic discs and/or bars \citep{Debattista1998,Errani2017},
dynamical friction from in-falling clumps \citep{El-Zant2001},
and mergers of substructures within or close to the halo \citep{Dekel2003}.

Although there is strong evidence that baryons have an impact on the dark matter haloes, it is still under debate
how the halo properties evolve with galaxy assembly history.
Furthermore, most of the reports focus only on the properties at $z=0$.
Thus, the investigation of the cosmic evolution of the dark matter haloes together
with the galaxy formation can help to disentangle the influence of baryons onto the dark matter haloes.

There are previous works that recently discuss this aspect. 
\citet{Chan2015} explore the dark matter profiles in the mass range of $\sim 10^9 - 10^{11} {\rm M}_{\odot}$
from the FIRE suite hydrodynamical simulations between $z=0-2$.
Their results suggest that the inner part of the dark matter density profiles change according to the halo mass.
Haloes with masses in the range of $10^{10} - 10^{11} {\rm M}_{\odot}$ have shallow profiles, while the 
baryonic contraction is significant for the haloes with masses above $5\times10^{11}~{\rm M}_{\odot}$.
Furthermore, they show that the inner regions of the dark matter haloes are governed by stellar mass and dark matter, and cores form typically close to $z\sim 0$. 
Also \citet{Zemp2012}, using hydrodynamical simulations with different prescriptions for the sub-grid physics of baryons, investigate 
the properties of haloes with masses above 10$^{10}~{\rm M}_{\odot}$ at $z\geq2$.
They find that dark matter haloes contract in response to baryon dissipation, producing approximately 
isothermal profiles in the inner 10\% of 
the virial radius. Moreover, they find that the specific angular momentum of dark matter is conserved 
in time in the region dominated by baryons, since
the stellar content stabilises the gravitational potential. 
\citet{Tollet2016} using a sample of dark matter haloes in the range of $10^{10} -10^{12} {\rm M}_{\odot}$,
find that the inner dark matter density profile slope depends on the stellar to halo mass ratio 
at all redshifts. This result is in agreement with \citet{DiCintio2014b} which investigate this dependence at $z=0$.

While previous reports explore the evolution of dark matter profiles in a variety of halo masses, in this
work we focus on how the physics of baryons affects the evolution of Milky Way mass-sized
dark matter haloes. This work extends the analysis performed by \citet{Tissera2010} at $z=0$.
We analyse four haloes from the Aquarius Project  \citep{Springel2008, Vogelsberger2009}
between $z$=0--4, and compare the cold dark matter only run with one of the galaxy formation models implemented 
in these haloes \citep[the hydrodynamical model from][]{Scannapieco2009}.
Understanding the effect of baryons on the dark matter density profile, and in particular in the inner regions
of Milky Way mass-sized haloes, is essential for making predictions of indirect dark matter detection
such as the gamma-ray signal from dark 
matter self-annihilation \citep{Abazajian2014,Schaller2016b}.
We acknowledge the fact that the Aquarius haloes of \citet{Scannapieco2009} overproduced stars, 
resulting in galaxies with large bulges and small discs.
Nevertheless, these haloes can provide information on the physical processes intervening.
To assess if our conclusions might be affected by this aspect, we also analysed
a version of one of the haloes run using AREPO code \citep{Marinacci2014}.
We show that the main results remain valid.

This paper is organised as follows. We describe briefly the numerical simulations and discuss the convergence of the
results in \S~\ref{sec:simulation}. In \S~\ref{sec:results} we investigate the evolution of different properties such as the dark matter density profiles,
the mass accretion history, the halo shape, and the specific angular momentum.
We contrast our findings between the dark matter only run of each halo and the simulation 
containing baryons. 
We summarise our main results in \S~\ref{sec:conclusions}.

\section{Simulation}
\label{sec:simulation}

The haloes of Aquarius project \citep{Springel2008, Vogelsberger2009} were run originally with dark matter particles only (DMO run), 
but subsequently, different methods were implemented to model the physics of baryons
\citep{Scannapieco2009,Cooper2010,Marinacci2014}. In this work we compare the results from the DMO with those
from the hydrodynamical run of \citet[SPH run]{Scannapieco2009}.
Both DMO and SPH simulations were run from $z=127$ to $z=0$ 
using versions of {\sc GADGET-3}, an update of {\sc GADGET-2} 
\citep{Springel2001,Springel2005} optimised for massively
parallel simulation of highly in homogeneous systems.
The initial conditions were settled depending on the mass resolution, 
and for the SPH runs the mass of dark matter particles was reduced accordingly to the cosmological parameters adopted
\citep[for further details see,][]{Springel2008, Scannapieco2009}.
The cosmological parameters adopted for SPH and DMO runs are
$\Omega_{\rm m} = 0.25$, $\Omega_{\rm b} = 0.04$,
$\Omega_{\Lambda} = 0.75$, $\sigma_{8} = 0.9$, $n_{s} = 1$, ${\rm H}_{0} = 100~h {\rm km~s^{-1} Mpc^{-1}}$ with $h=0.73$.
These cosmological parameters are the same adopted in the Millennium Simulation \citep{Springel2005b}, which 
within the uncertainties, are consistent with the constraints derived from WMAP 1- and 5-year data analyses \citep{Spergel2003,Komatsu2009}.
The evolution of the structure at galactic scales are not significantly affected by small variations in the cosmological parameters.

The DMO haloes were run with different resolutions as part of the Aquarius Project.
\citet{Navarro2010} analyse the properties of the DMO haloes, showing that they converge numerically  
in all the resolution levels studied, between level 1 to 5.
So, it is possible to reliably use level-2 of the DMO to compare with the SPH runs at level-5 as also explained in \citet{Tissera2010}.
The haloes that include baryonic processes were run at a lower resolution approximately 200 times lower than the DMO, named as level-5.
We also analyse the results of a lower resolution run of level-6 for one of the haloes to evaluate resolution effects.
In general, the convention adopted to label each halo is with a letter and the resolution level.

From the set of six haloes analysed by \citet{Tissera2010} at $z=0$, we studied 
the evolution of four of them selected to not to have any major event which left 
distinct features in the dark matter distribution. This is the case for Aq-B-5 and Aq-F-5.
Therefore the haloes from the hydrodynamical simulation analysed are Aq-A-5, Aq-C-5, Aq-D-5, Aq-E-5. We also analysed a lower resolution run Aq-E-6
to assess numerical issues.
The gravitational softening of SPH runs are in the range of $\epsilon_{\rm G, SPH} = 0.5 - 1 \,h^{-1}\,{\rm kpc}$. 
The dark matter particles in the SPH simulation are $\sim 1-2 \times 10^{6} h^{-1} {\rm M}_{\odot}$, while gas particles have initially
masses of $\sim 2-5 \times 10^{5} h^{-1}\,{\rm M}_{\odot}$. 
The main properties of the dark matter haloes including the number of dark matter
particles within the virial radius are presented in Table~\ref{tab:fits}.
Following the same convention, and according to the resolution, the DMO haloes are generally named as Aq-A-2, Aq-C-2, Aq-D-2 and Aq-E-2.
The gravitational softening of the DMO runs is $\epsilon_{\rm G, DMO} = 0.048  \,h^{-1}\,{\rm kpc}$, while
the mass of dark matter particles are $0.7-1 \times 10^{4} h^{-1}\,{\rm M}_{\odot}$.
All the results of the DMO haloes presented are scaled by the cosmic baryon fraction, f$_{\rm bar} = \Omega_{\rm b}/\Omega_{\rm m}$.

In order to avoid confusion and make the nomenclature simpler for the reader, in this work we 
name the haloes of the SPH run as Aq-A-SPH, Aq-C-SPH, Aq-D-SPH and Aq-E-SPH,
while those from the DMO run are refereed as Aq-A-DMO, Aq-C-DMO, Aq-D-DMO and Aq-E-DMO.

We study the evolution of the dark matter haloes in the range of $z \sim 4$ to $z \sim 0$ in both
DMO and hydro simulations.
In the following section, we describe briefly the main properties
of the baryonic model implemented in the hydrodynamical simulation.

\subsection*{The hydrodynamical model}

The simulations were run by \citet{Scannapieco2009} using a version of {\sc P-GADGET-3}  that include 
a multiphase model for the gas component with metal-dependent cooling, star formation and phase-dependent treatments of
supernova (SN) feedback and chemical enrichment \citep[see][for further details]{Scannapieco2005,Scannapieco2006}.

Gas particles are eligible to form stars according to their density (must be denser than ${\rm n_{H,th}} = 0.04 {\rm cm}^{-3}$) 
and if they lie in a convergent flow. Stellar particles are created stochastically from those gas particles
that fulfil the aforementioned conditions, and a maximum of two stellar particles are allow to form from each gas particle
\citep[for further details see][]{Scannapieco2006,Scannapieco2009}.
The chemical model implemented follows the chemical enrichment by supernovae type II (SNII) and
type Ia (SNIa) from \citet{Mosconi2001}. 
The SN feedback model has been proven to be effective at reproducing 
the observed phenomenology of star formation and wind generation in quiescent
and starburst galaxies. Hence, the algorithm implemented in the SPH runs have shown to be a powerful
tool to investigate galaxy formation within a cosmological context \citep[][]{Scannapieco2009}.

Haloes are identified at their virial radius $R_{\rm 200}(z)$, defined as the radius which encloses a 
density equal to $\sim$200 times the critical density at certain redshift $z$. The halo mass M$_{{\rm 200},z}$, is defined as the mass enclosed
within $R_{{\rm 200},z}$. We summarise the main properties of the haloes in Table~\ref{tab:fits}.

The dark matter haloes from the SPH run were previously studied in \citet{Tissera2010} at $z=0$. 
It is already established that Aquarius haloes have diverse assembly histories which produce a variety
of star formation histories and structures. The properties of the central galaxies within the Aquarius haloes
are analysed in \citet{Scannapieco2009}, finding that most of them present centrifugally supported discs
composed of approximately a fifth of the total stellar mass (with exception of Aq-F-5 which is not studied in the present work).
The star formation rate has shown to be different for each central galaxy, as well as for spheroidal and disc components.
Therefore, we expect that the different assembly histories will produce different evolution for the dark matter halo shapes.
The present work is an extension of that of \citet{Tissera2010} in order to investigate 
the cosmic evolution of dark matter haloes and the role that baryons
played on them.

We acknowledge the fact that these Aquarius haloes have
overproduced stars at $z=0$ compared with abundance matching results
\citep[e.g.,][]{Guo2010,Behroozi2018,Moster2018} as already discussed by
\citet[][]{Aumer2013}.
To assess the robustness of our analysis with respect to this issue, 
we also analyse a version of Aq-C performed with the moving mesh code {\sc arepo} 
\citep[][Aq-C-M14, run at resolution level-4]{Marinacci2014}.  
The sub-grid model accounts for star formation, chemical enrichment, stellar feedback using
a kinetic wind scheme, metal-line cooling, and quasar- and radio- mode
feedback from AGN.

This run has produced galaxies with disc-dominated systems and stellar mass to halo mass ratio in agreement with
the results from abundance matching techniques at $z=0$.
Additionally, the comparison of Aq-C-SPH with Aq-C-M14 will allow us to
assess effects of numerical resolution. 

\begin{figure}
 \centering
  \includegraphics[width=0.5\textwidth]{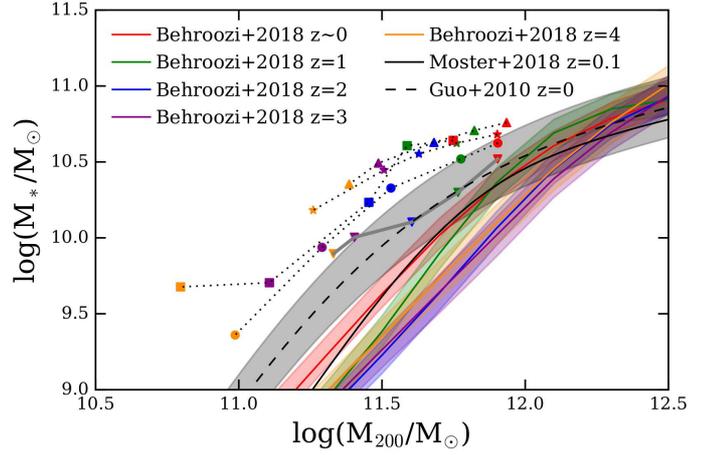}
 \caption{Stellar mass to halo mass relation for the five haloes
   studied: Aq-A-SPH (stars), Aq-C-SPH (triangles up), Aq-D-SPH (circles), Aq-E-SPH (squares) and Aq-C-M14 (triangles down).   
 The results for the five haloes are plotted where color code represents the different redshifts
 ($z=0$ red, $z=1$ green, $z=2$  blue, $z=3$ purple, and $z=4$  orange). We follow the evolution of the SPH haloes with dotted lines. 
 The grey filled line connects the different refshifts for Aq-C-M14.  
 We compared our findings with the results of abundance matching technique from \citet{Guo2010}, \citet{Behroozi2018}, and \citet{Moster2018} (see labels in the figure).}
  \label{fig:SMHM_Comparison}
 \end{figure}

In Fig.~\ref{fig:SMHM_Comparison} we show the cosmic evolution from $z=0$ to $z=4$ of
the stellar mass to halo mass relation for the five haloes compared with 
abundance matching results \citep{Guo2010,Behroozi2018,Moster2018}.
At $z=0$, three of the galaxies are in agreement with the results from
\citet[][within $\sim 1 \sigma$ error]{Guo2010}: Aq-A-SPH, Aq-D-SPH
and Aq-C-M14. The remaining two haloes contain an excess of stellar mass
although they are within $ 2 \sigma$ error
(Aq-C-SPH and Aq-E-SPH). 
At higher redshifts, the all simulated haloes have a higher discrepancy compared with abundance matching techniques, showing
they form a higher fraction of stars. As can be seen from
Fig.~\ref{fig:SMHM_Comparison} all haloes show a similar increase of the
stellar mass to halo mass with redsfhit.  Interestingly, Aq-C-M14  that is well consistent with the abundance
matching predictions at $z=0$ does not reproduce the cosmic evolution
of predicted values.
Hence, there are still unclear aspects regarding the regulation of the
star formation activity as a function of redshift and hence, on the
impact of the baryons on the dark matter distribution.
We will use these set of simulated haloes to analyse the
response of the dark matter to galaxy formation as the they are
assembled baring in mind the caveats. The comparison of haloes with
different history of assembly run with the same subgrid physics  as
well as the comparison between runs with the same initial conditions
and different subgrid physics provide a route to contribute to
understand this problem.

\subsection*{Spatial resolution and convergence considerations}\label{sec:resolution}

In order to study the inner region of the dark matter haloes, it is important to analyse 
the minimum radius above which the results
obtained are robust. This radius is named as the convergence radius.
For DMO simulations, \citet{Power2003} find that the converge radius R$_{\rm P03}$ should not
exceed the radius where the two body relaxation
time is shorter than the age of the Universe. 
However, there is no a straightforward definition for hydrodynamical simulations.
Nonetheless, the converge criterion of \citet{Power2003} might be used as reference in hydrodynamical simulations,
considering it is a conservative representation of the convergence radius when baryons 
are included in simulations \citep{DiCintio2014b,DiCintio2014,Schaller2015}.
Adopting the R$_{\rm P03}$ estimation for SPH runs, we find that the values are around 1.50--2.50~$h^{-1}$~kpc at $z=0$
which represent roughly one percent of the virial radius. This value represents approximately three times the gravitational softening implemented in the SPH haloes at $z=0$.
This result is similar at high redshift, finding that the convergence radius is below to 5 percent of the virial radius at each redshift. 
For the dark matter haloes of the DMO runs, the convergence radius is lower since the resolution is much higher than that
of the SPH runs.

Therefore, in order to study the evolution of the dark matter profiles and the changes in the inner region, we consider the convergence
radius of \citet{Power2003} (i.e., $\sim$ 1 percent) as the minimum radius where we can analyse the SPH haloes.
In this work, we adopt different limits according to the halo properties we want to study and considering the convergence radius obtained 
for these simulations. For instance, we fit the dark matter halo profiles between three times the gravitational softening and the virial radius
at each redshift in order to fulfil this condition. 
It is also important to be careful with the number of particles on each radius bin for the
dark matter profile. We describe the adopted criterion for SPH and DMO haloes in \S~\ref{sec:results-DMProf}.

Considering the convergence radius, and in order to make a robust analysis, we define the inner region of the haloes as the region within
5 percent of $R_{200}$.

\section{Results}
\label{sec:results}

\subsection{Dark matter density profiles at $z=0$}\label{sec:results-DMProf}

\begin{figure*}
  \includegraphics[width=0.5\textwidth]{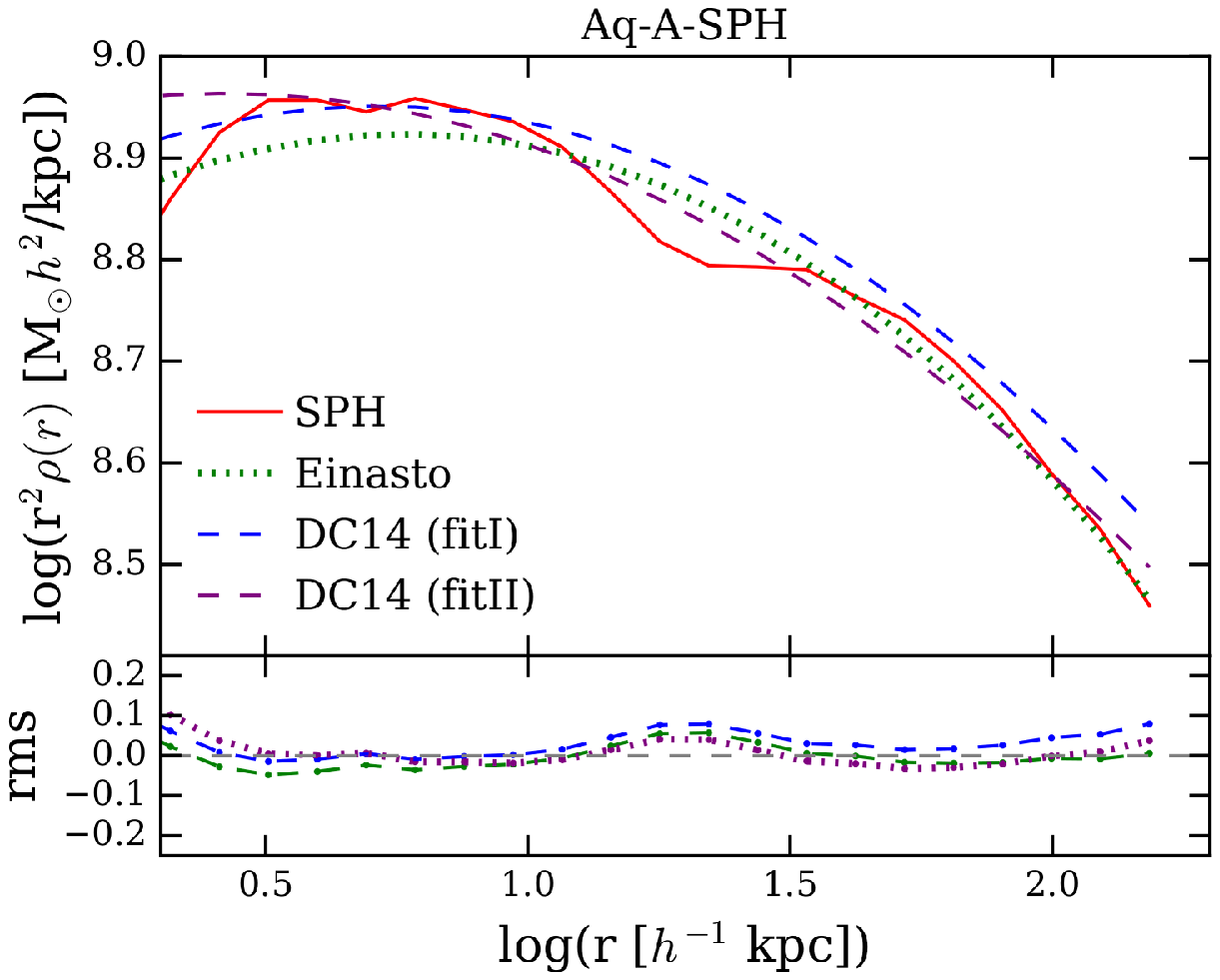}
  \includegraphics[width=0.5\textwidth]{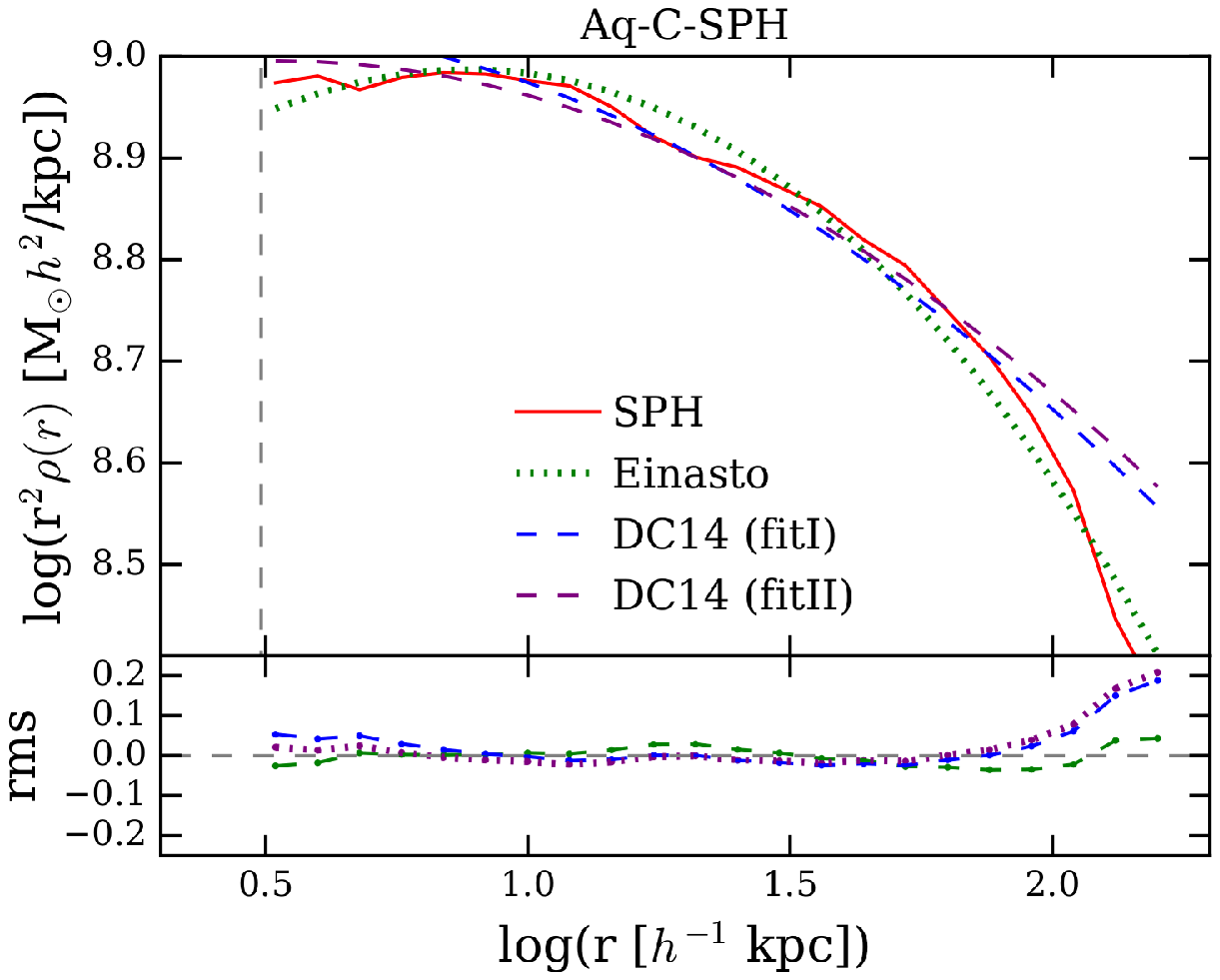}
  \includegraphics[width=0.5\textwidth]{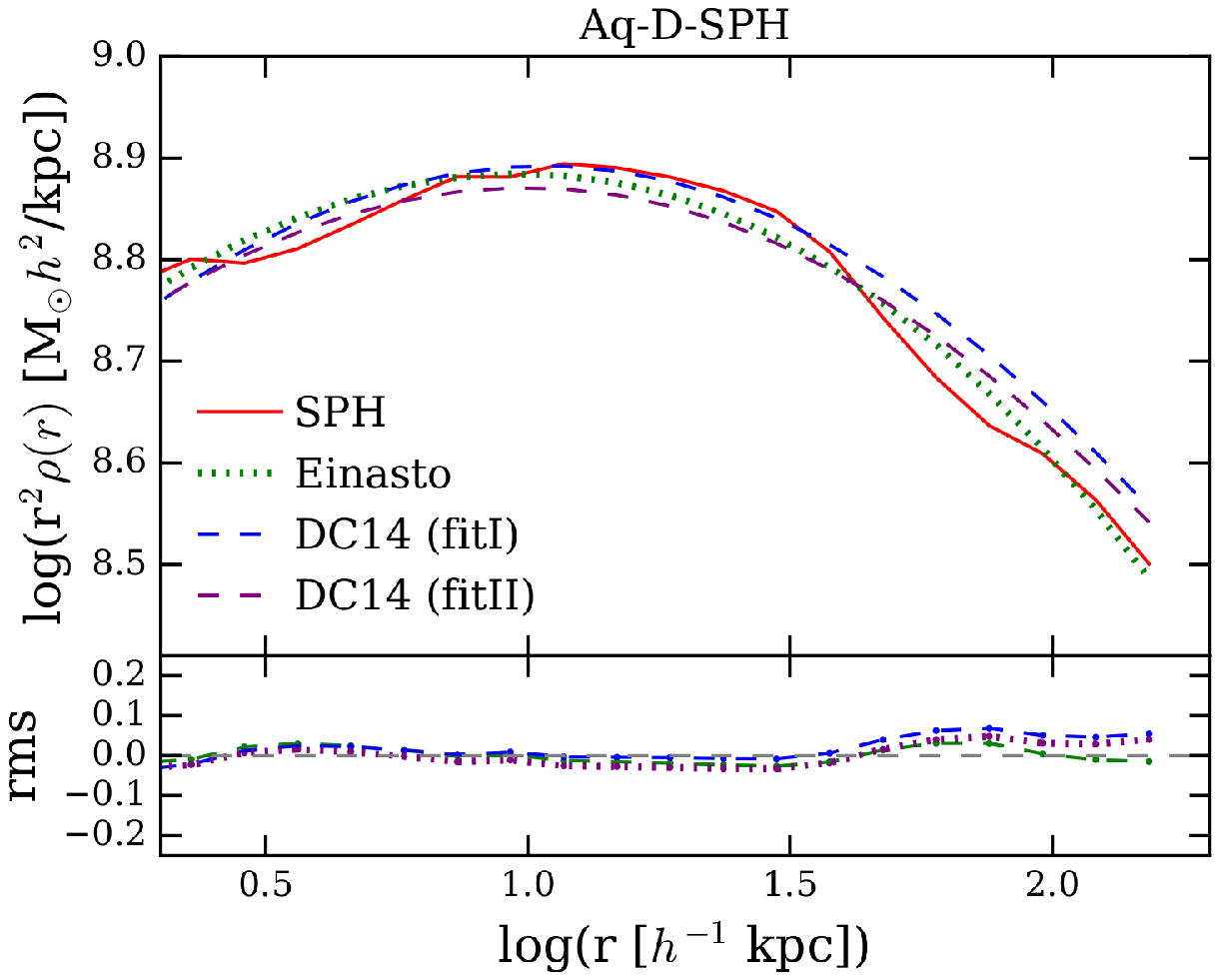}
  \includegraphics[width=0.5\textwidth]{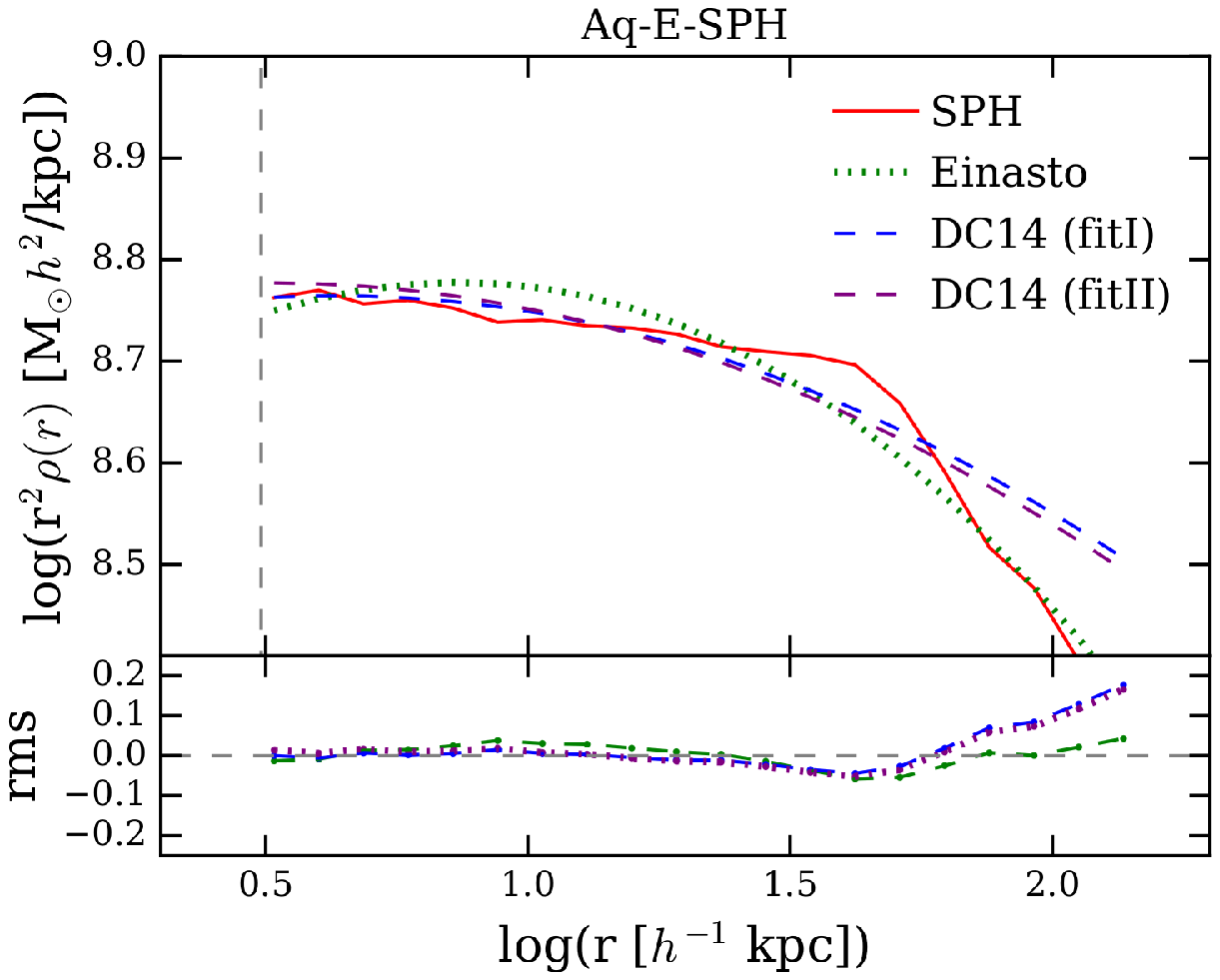}
 \caption{Comparison of the spherically averaged dark matter profiles from the SPH run at $z\sim0$ with the fits of 
 Einasto profile (green dotted lines) and \citet{DiCintio2014} (DC14) profile.
 In the case of DC14, we adopt two methods: the first assumes the concentration parameter DMO as a free parameter
 (named as DC14 fitI, blue dashed lines); 
 and the second method considers both the concentration parameter and the halo mass as free parameters (named as DC14 fitII, dashed purple line). 
 Grey dashed line in the profiles Aq-C-SPH and Aq-E-SPH5 represent three times the gravitational softening for these haloes. 
 We also include the rms for each fit in the lower panels.}
  \label{fig:fits-profile}
 \end{figure*}
 
 \begin{figure*}
  \includegraphics[width=0.5\textwidth]{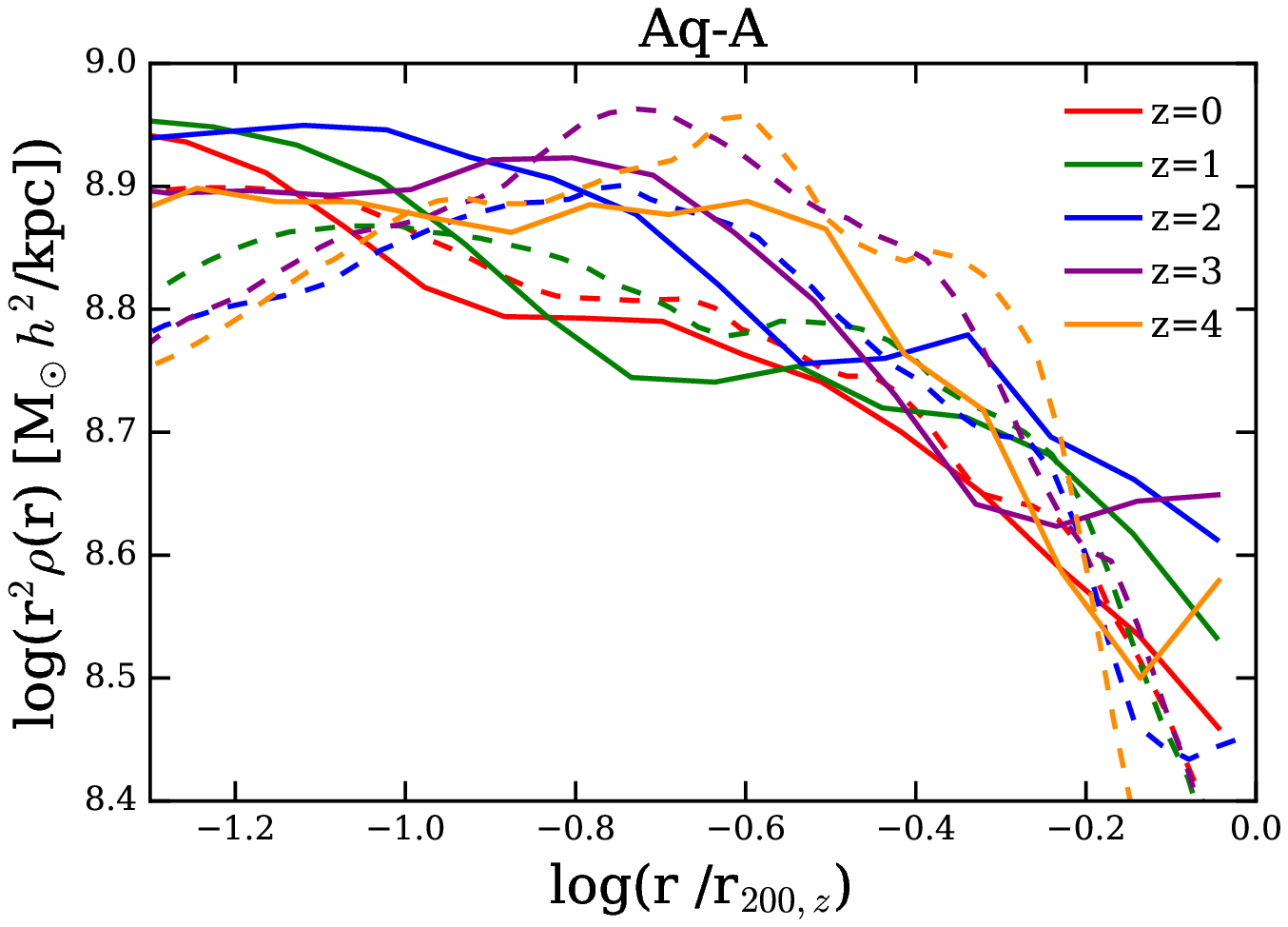}
  \includegraphics[width=0.5\textwidth]{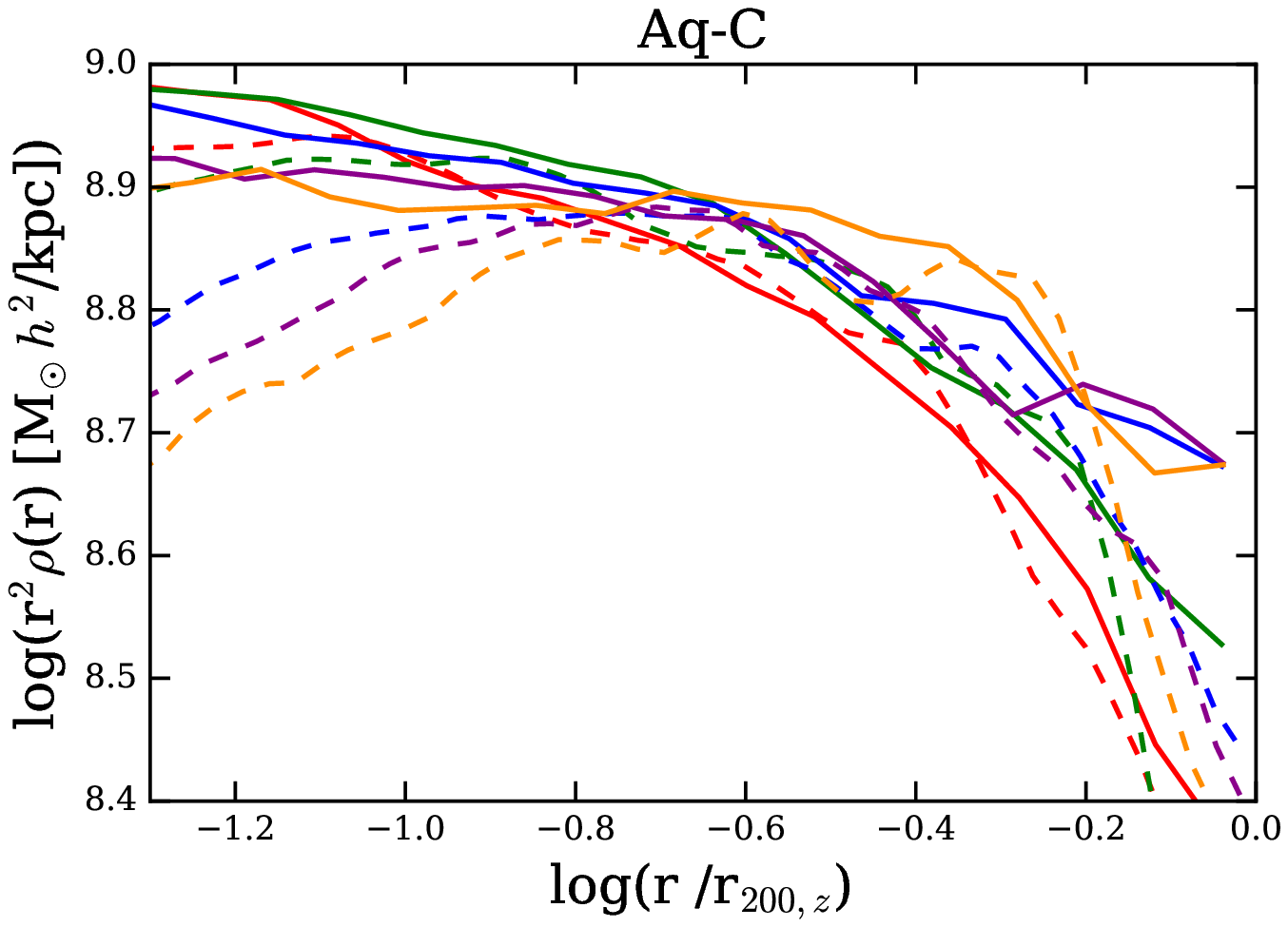}
  \includegraphics[width=0.5\textwidth]{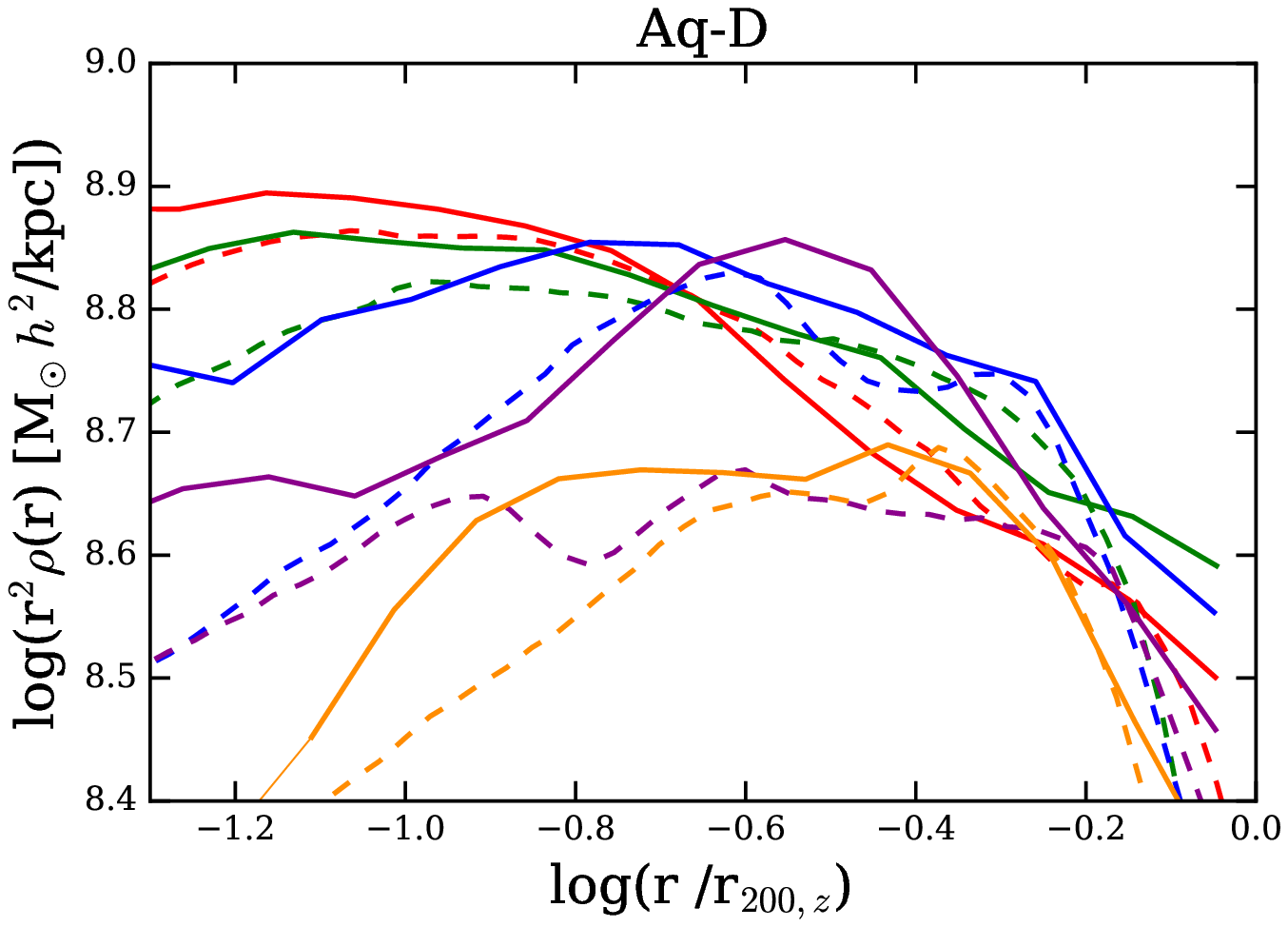}
  \includegraphics[width=0.5\textwidth]{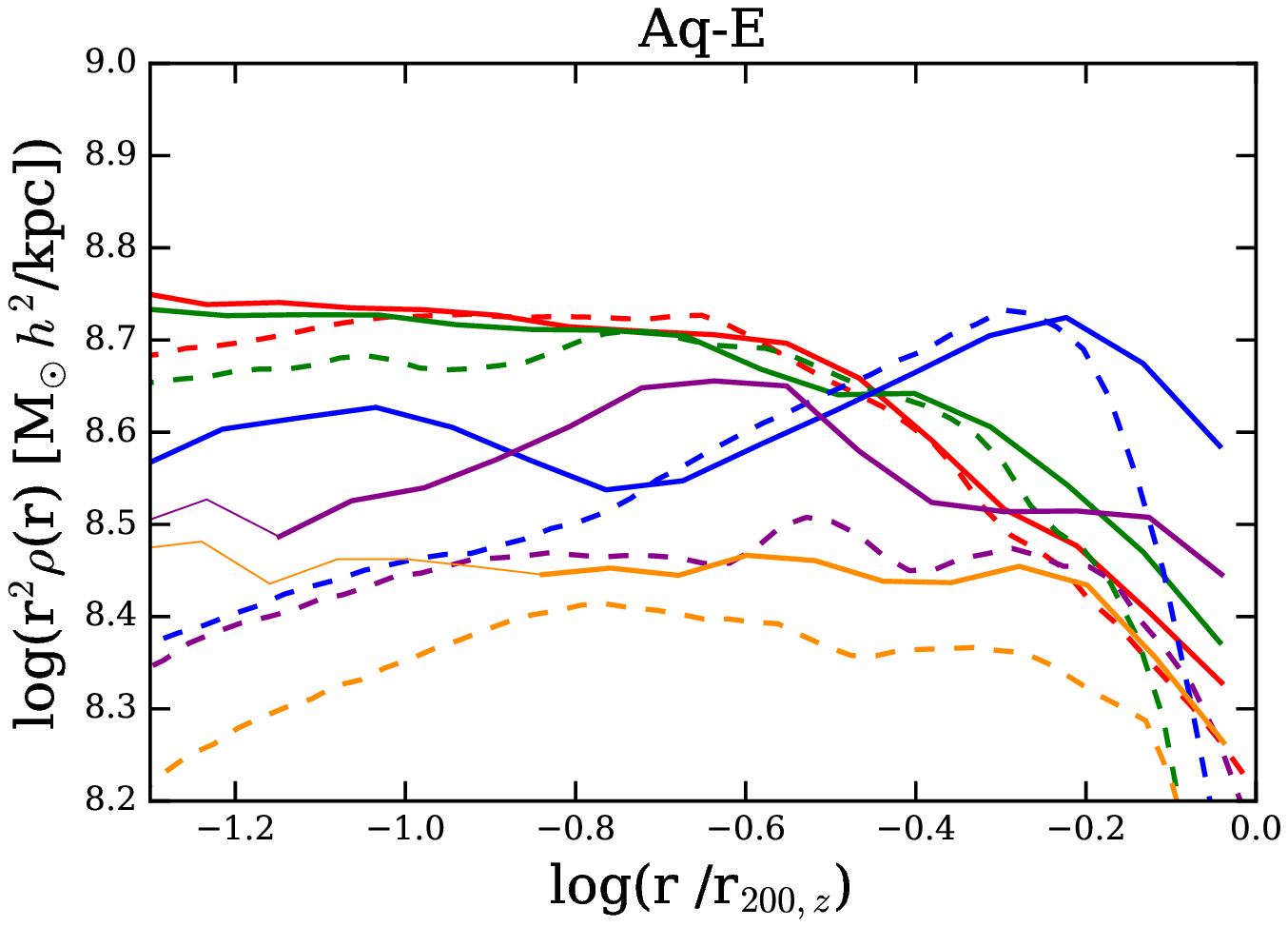}
 \caption{Dark matter density profiles at redshift $z = 0, 1, 2, 3$ and 4 for SPH (filled lines)
 and DMO (dashed lines) simulations. 
  For SPH haloes, thin lines show the bins that contain less than 2000 particles.
  The radius range presented is in agreement with the convergence criteria of  \citet{Power2003}.}
 \label{fig:DM-profiles2}
\end{figure*}

\begin{figure}
 \centering
  \includegraphics[width=0.5\textwidth]{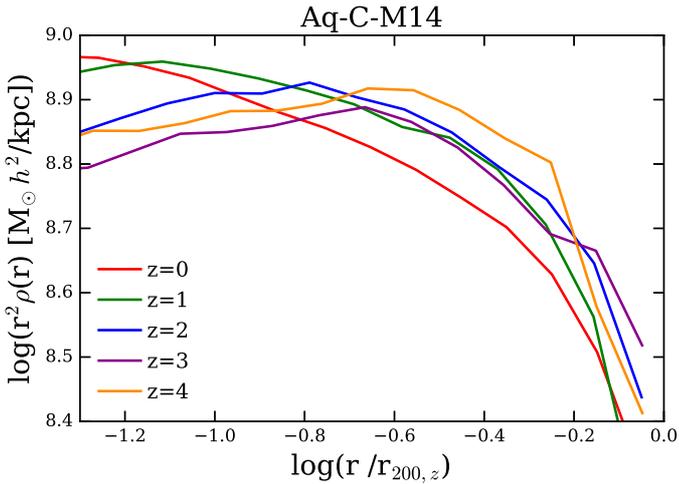}
 \caption{Same than Fig.~\ref{fig:DM-profiles2} for the halo Aq-C-M14.}
 \label{fig:DM-profiles2M14}
\end{figure}
 
We analyse the dark matter profiles of the SPH runs at $z=0$.
For this, we implement two different models, the one proposed by \citet{DiCintio2014}, and the Einasto 
profile \citep{Einasto1965} also shown in \citet{Tissera2010}.

The profile proposed by \citet[][DC14]{DiCintio2014}, is based on a double power-law model
with three parameters ($\alpha$, $\beta$, $\gamma$) fitted depending on the star formation efficiency of the galaxy 
(i.e., the relation between stellar mass and dark matter halo mass).  
The model also includes $\rho_s$ the scale density, and $r_s$ the scale radius,
which depend on the concentration parameter of the DMO halo \citep[see][for further details]{DiCintio2014}.
Hence, the functional form of \citet{DiCintio2014} profile is:

\begin{equation}\label{eq:dc14}
 \rho(r) = \frac{\rho_{s}}{\big( \frac{r}{r_{s}} \big)^{\gamma} \big[1 + \big(\frac{r}{r_{s}} \big)^{\alpha}\big]^\frac{(\beta -\gamma)}{\alpha} }.
\end{equation}

In DC14, the parameters that depend on the star formation efficiency are modelled by two functions. 
The outer slope $\beta$ is fitted with a quadratic function, while
the inner slope $\gamma$ and the transition parameter $\alpha$ are fitted by power-law functions \citep[see][eq.~3]{DiCintio2014}.
Since $\alpha$, $\beta$, and $\gamma$ parameters are constrained trough the star formation efficiency, 
the DC14 profile is left with 2 free-parameters, alike the NFW one.
In order to get the free parameters of the functional form, DC14 select a sample of ten galaxies with different stellar masses 
and five different initial conditions, with efficiencies in the range of $-4.1 < \log ({\rm M}_{*} /{\rm M}_{200} ) < -1.3$.
This profile turns towards NFW profile
for star formation efficiencies around $\log({\rm M}_{*}/{\rm M}_{200}) = -1.5$ and above, since outflows
are not efficient enough to pull out the mass from the inner regions,
getting values similar to a NFW profile in this case ($\alpha = 1$, $\beta = 3$ and $\gamma = 1$).

The star formation efficiency of the central galaxies from the SPH haloes are $\log({\rm M}_{*}/{\rm M}_{200}) =$ -1.22 (Aq-A-SPH),
-1.18 (Aq-C-SPH), -1.28 (Aq-D-SPH) and -1.11 (Aq-E-SPH).
If we compare the efficiencies from the SPH run with the range studied by DC14, 
we find that the Aquarius haloes are slightly in excess the range studied by around $\sim 0.1-0.2~{\rm dex}$. 
Hence, we implement the expressions from DC14 for $\alpha$, $\beta$, and $\gamma$, and leave the concentration and halo mass
as free parameters to be determined through fitting procedures
\citep[see also][]{Katz2017}.
We fit Eq.~\ref{eq:dc14} to each halo 
between three times the gravitational softening to the virial radius adopting:
the concentration as the free parameter (DC14 fitI);
and also adopting the concentration and halo mass as free parameters (DC14 fitII).
We present the parameters from DC14 fitI and fitII in Table~\ref{tab:DC_par}, and the profiles obtained in Fig.~\ref{fig:fits-profile}.
For each SPH halo we compute the root-mean-square deviation (rms), presented in the lower panel of each halo.

The concentration parameters from DC14 (see Table~~\ref{tab:DC_par}) can be directly compared to 
\citet{Ludlow2014}, since the scale radius $r_s$ has been converted into $r_{-2}$ using Eq.~2 in \citet{DiCintio2014}.
We find that the concentration parameters obtained present values in agreement with those from \citet{Ludlow2014}.

 \begin{table} 
\begin{center}
\caption{Parameters obtained from fitting DC14 profile fitI and fitII (see the text for details).
The halo masses obtained from fit II, M$_{200}^{\rm DC14,II}$, are in units of $10^{12}~h^{-1}$~M$_\odot$.
We also include the $\chi^{2}$ test of the model for each halo.} 
\label{tab:DC_par}
\begin{tabular}{lcccc}
\hline \hline
                     & Aq-A-SPH  &  Aq-C-SPH  &  Aq-D-SPH &  Aq-E-SPH  \\ 
\hline
Fit I \\
$\alpha$             &  0.40     &   0.30     &    0.53   &   0.24    \\
$\beta$              &  2.98     &   3.01     &    2.94   &   3.03    \\
$\gamma$             &  1.28     &   1.33     &    1.22   &   1.35    \\
c$_{\rm DM}$         &  10.59    &   21.72    &    5.86   &   9.61    \\
$\chi^{2}$            &  0.008    &   0.017    &    0.004  &   0.015   \\
\hline
Fit II &&&& \\
$\alpha$             &  0.32     &   0.31     &    0.49   &   0.22   \\
$\beta$              &  3.00     &   3.01     &    2.95   &   3.04   \\
$\gamma$             &  1.32     &   1.32     &    1.24   &   1.36   \\
c$_{\rm DM}$         &  19.13    &   15.39    &    6.09   &   12.57  \\
M$_{200}^{\rm DC14,II}$  & 1.0   &   1.20     &    1.05   &   0.78   \\
$\chi^{2}$            &  0.007    &   0.020    &    0.002  &   0.013  \\
\hline
\end{tabular}
\end{center}
\end{table}

We find that both fitI and fitII provide good agreement in the inner regions of the haloes, which is represented by a steep
inner slope of $\gamma \sim 1.2-1.4$ (see Table~\ref{tab:DC_par}) indicating contracted haloes at the centre.

However the model seems to show some discrepancies in the outskirts of the SPH haloes.
These discrepancies might be due to different aspects, such as the merger history of some Aquarius haloes. 
Indeed, the outer slope of some of the Aquarius haloes might be still  affected at $z=0$ by some of the late mergers,
as a result of their outer slope being steeper than the canonical slope of -3 usually found in DM haloes.
This is in particular the case of Aq-E-SPH which presents a peculiar dark matter density profile 
\citep[see also][for a discussion regarding Aq-E-DMO]{Navarro2010}.
Another aspect is that the dark matter haloes implemented in DC14 are mainly in
the range of $1.3\times10^{10}$-$9.9\times10^{11}$~$h^{-1}$~M$_{\odot}$, while Aquarius are more massive haloes,
although this should not  be an issue when the outskirts of dark matter halos is studied.

We also fit the dark matter profiles with Einasto profile function at $z=0$ between three times the gravitational softening and 
the virial radius  \citep[also shown in][but fitted in a different range]{Tissera2010}.
This profile describes the dark matter density using three free parameters $\alpha$, $r_{-2}$ and $\rho_{-2}$, 
where the second and third parameters are the isothermal radius and density, when the logarithmic slope is -2.
The Einasto profile functional form is:

\begin{equation}
 \rho(r) = \rho_{-2} \exp \big[\frac{-2}{\alpha} \big(\big(\frac{r}{r_{-2}}\big)^{\alpha} - 1\big) \big].
\end{equation}

In Fig.~\ref{fig:fits-profile} we show the result obtained for the SPH haloes at $z=0$, and the values obtained for the parameters
are presented in Table~\ref{tab:fits}.
We find that Einasto provides a good fit for all the analysed haloes in agreement with previous results \citep{Tissera2010}.
From the comparison of the $\chi^{2}$ values, we find that Einasto profile
reproduces slightly better the SPH dark matter haloes profiles than DC14.
This result is also expected since the Einasto contains an extra free parameter.

In Table~\ref{tab:mass} we show the comparison of the halo mass of each SPH halo with the masses inferred from the fits
of Einasto and DC14 fitII at $z=0$. We find that the masses computed are in good agreement with the simulated haloes, with deviations between 
one to ten percent the original halo mass. 

 \begin{table} 
\begin{center}
\caption{Dark matter halo mass at $z=0$. We compare the mass of the SPH runs (M$_{200}^{sim}$) with those inferred from
fitting Einasto profile (M$_{200}^{E}$) and \citet{DiCintio2014} fitII (M$_{200}^{\rm DC14,II}$). See Sec.~\ref{sec:results-DMProf} for further details.} 
\label{tab:mass}
\begin{tabular}{lcccc}
\hline \hline
Name                 & M$_{200}^{sim}$                   &  M$_{200}^{E}$ &   M$_{200}^{\rm DC14,II}$  \\ 
                     &  $\times 10^{12} h^{-1} {\rm M}_{\odot}$ &  $\times 10^{12} h^{-1} {\rm M}_{\odot}$  &  $\times 10^{12} h^{-1} {\rm M}_{\odot}$ \\
\hline 
Aq-A-SPH               &    1.10                            &  0.99           &       1.00                \\
Aq-C-SPH               &    1.18                            &  1.07           &       1.20                 \\
Aq-D-SPH               &    1.09                            &  1.01           &       1.05                  \\       
Aq-E-SPH               &    0.77                            &  0.69           &       0.78                 \\
\hline
\end{tabular}
\end{center}
\end{table}

\subsection{Evolution of dark matter density profiles with redshift}\label{sec:results-DMProfEv}

In this section, we follow the evolution of the dark matter halo profiles and compare the results obtained from DMO and SPH runs. 
We show that the evolution is different between DMO and SPH haloes, suggesting that these differences lie on the 
impact that baryons have onto the SPH dark matter haloes.  
Furthermore, the different SPH haloes studied also show particular trends.

We do not implement the model from DC14 at higher redshifts into the SPH haloes
since the parameters that depend on the star formation efficiency were adjusted to reproduce the halo population at $z=0$. 
Only the evolution of the inner slope has been modelled in the past \citep{Chan2015,Tollet2016}.
Different reports have shown that the star formation efficiency depend on redshift
according to different processes such as star formation and merger history of the haloes \citep[e.g.,][]{Behroozi2013}.
Hence, an alternative model that accounts for the evolution of the star formation efficiency with redshift 
would be essential to develop in the future.  

We use the Einasto profile to model the dark matter haloes at $z=0, 1, 2, 3$ and 4, to quantify the differences as a function of redshift.
We fit the free parameters between three times the gravitational softening and the virial radius at each redshift, R$_{200,z}$.
In Fig.~\ref{fig:DM-profiles2} we present the dark matter density profiles for both DMO and SPH runs at redshift $z=0, 1, 2, 3$ and 4.
We compute the dark matter density profiles of DMO haloes between three times the gravitational softening to ${\rm R}_{\rm 200,z}$, using 100 bins.
For the SPH haloes the resolution is lower, hence we use 22 bins in order to have more than 2000 particles 
in each radial bin. When the bin does not fulfil this condition the line of the 
profile is shown with a thin line.

All the DMO haloes show that the amplitude of the density profile increases as redshift decrease in the 
inner region below ten percent of the R$_{200,z}$, turning from a NFW profile at high redshift ($\rho(r) \propto r^{-1}$)
to isothermal profile ($\rho(r) \propto r^{-2}$). We find that the DMO profiles present similar characteristics in the outskirts. 
For the SPH runs, the DM halo profiles do not show the same evolution. In particular, haloes Aq-A-SPH and Aq-C-SPH appear
to be roughly stable and do not show critical evolution with redshift in the inner parts. 
More precisely, Aq-A-SPH present some slight variations while Aq-C-SPH is more stable in time. 
This is not the case for Aq-D-SPH and Aq-E-SPH which show an increase in the inner region, similar to what
we find in the DMO haloes. We note that the inner region of Aq-E-SPH
at $z\gtrsim3$ contain less than 2000 particles per bin.
Hence the analysis of Aq-E-SPH must be taken with caution above $z\sim 3$.

The comparison of the inner radii of SPH and their counterparts DMO show that SPH haloes are more concentrated
as found in other several works \citep[e.g.,][]{Tissera2010,Schaller2016,Zhu2016}.
Furthermore, we find that the SPH dark matter profiles in the inner region below  
ten percent of $R_{200,z}$ and between $z = 0 - 2$, are described by isothermal profiles.
This trend is more clear for Aq-A-SPH, Aq-C-SPH, and Aq-E-SPH.

It is also interesting to point out that both, Aq-A-SPH and Aq-C-SPH, end up being more concentrated in the inner regions at $z=0$.
In Table~\ref{tab:fits} we present the concentration parameters for the redshifts studied, defined as $c_{\rm SPH}=R_{200}/r_{-2}$
where $r_{-2}$ is obtained from fitting Einasto profile.
At $z=0$ we find that the concentration parameter is higher for Aq-A-SPH and Aq-C-SPH compared with Aq-D-SPH and Aq-E-SPH. This difference
might be due to different halo assembly histories together with baryon accretion in the haloes.
The concentration parameter shows to be dependent on the redshift and halo mass as expected \citep{Ludlow2014},
showing a tendency to decrease as redshift increase.

In Fig.~\ref{fig:DM-profiles2M14}, we present the dark matter density profile of Aq-C-M14 between $z=0-4$.
Aq-C-M14 show a mild evolution in the inner region, and a stable density profile between $z=0-1$, 
in contrast with Aq-C-SPH which presents a stable profile in almost all the redshift range studied.
The differences found between these two runs can be explained mainly due to the differences 
of sub-grid physics. Also the impact of the moving-mesh technique might improve
the mixing between cold and hot gas \citep[see e.g.,][]{Schaller2015}.

In the following sections we discuss the principal
differences we find in the evolution of the dark matter haloes and the effect that galaxy formation has on them through cosmic time.

\begin{table*}
 \caption{Properties of the dark matter haloes from the SPH run at $z = 0,1,2,3$ and 4: virial radius R$_{200,z}$, virial mass M$_{\rm 200,z}$, and number of
 dark matter particles within the virial radius N$_{200}^{DM}$.
 We present the parameters obtained from Einasto profile $\alpha$, $\log(\rho_{-2}$),  
 and the concentration parameter defined as $c_{\rm SPH}=R_{200,z}/r_{-2}$. We also show the $\chi^{2}$ of the fits.} \label{tab:fits}
\centering
 \begin{tabular}{lcccccccc}
\hline \hline
 Name                 & $\alpha$ & $\log(\rho_{-2}$)            &  $r_{-2}$  &  $\chi^{2}$       &  c$_{\rm SPH}$      & R$_{200,z}$ & M$_{200,z}$ & N$_{200}^{DM}$\\  
                      &          &  M$_{\sun}\,h^{2}$/kpc$^3$   & $h^{-1}$~kpc  &         &         & $h^{-1}$~kpc   & $10^{12} h^{-1}~{\rm M}_\odot $          &       \\    
 \hline 
 Aq-A-SPH               &          &                              &            &          &         &            &          &                                  \\
 $z = 0$	      &  0.09    &      7.39              &  5.87      &  0.003    &  28.86   &  169.42   &  1.10    &  527620        \\               
 $z  = 1$             &  0.06    &      8.13              &  2.53      &  0.003    &  43.25   &  109.43   &  0.79    &  372467       \\
 $z  = 2$             &  0.08    &      7.95              &  3.16      &  0.002    &  22.20   &   70.15   &  0.55    &  267611       \\
 $z  = 3$             &  0.12    &      7.88              &  3.32      &  0.007    &  13.47   &   44.74   &  0.36    &  164520       \\
 $z  = 4$             &  0.15    &      7.99              &  2.88      &  0.005    &  11.39   &   32.81   &  0.25    &  114884       \\
 \hline
 Aq-C-SPH             &          &                              &            &            &       &            &          &               \\
 $z =0$ 	      &  0.13    &      7.21              &  7.73      &  0.002     & 22.40    &  173.19    &  1.18    &  680301      \\                
 $z  = 1$             &  0.10    &      7.54              &  5.29      &  0.001     & 21.47    &  113.59    &  0.91    &  512148      \\
 $z  = 2$             &  0.07    &      7.96              &  3.14      &  0.001     & 23.28    &   73.10     &  0.66    &  365642    \\
 $z  = 3$             &  0.06    &      8.35              &  1.94      &  0.001     & 24.24    &   47.03     &  0.42    &  229907       \\
 $z  = 4$             &  0.08    &      8.01              &  2.81      &  0.002     & 12.96    &   36.42     &  0.33    &  181789     \\
 \hline
  Aq-D-SPH            &          &                             &            &            &       &  \\       
 $z  = 0$	      &  0.11    &      6.91              &  9.62      & 0.001     &  17.74    & 170.63    & 1.09     & 602250     \\                
 $z  = 1$             &  0.08    &      7.13              &  7.19      & 0.001     &  15.82    & 113.70    & 0.82     & 437386     \\
 $z  = 2$             &  0.05    &      7.48              &  4.62      & 0.007     &  14.65    & 67.69     & 0.45     & 265752     \\
 $z  = 3$             &  0.07    &      7.14              &  6.14      & 0.021     &  6.70     & 41.12     & 0.24     & 142447      \\
 $z  = 4$             &  0.18    &      6.88              &  7.32      & 0.022     &  3.73     & 27.29     & 0.13     & 72326      \\
 \hline
  Aq-E-SPH              &          &                             &            &       \\
 $z =0$	             &  0.10    &       7.04              &  7.39      & 0.003     &  20.29    & 149.94     &  0.77   &  538991   \\                
 $z =1$             &  0.07    &       7.55              &  4.02       & 0.007    &  23.63    &  95.00      &  0.53   &   359610  \\                
 $z =2$	             &  0.01    &       6.58              &  10.57     & 0.022     &  6.08     & 64.24      &  0.40   &   275840  \\                
 $z =3$	             &  0.11    &       6.92              &  6.83      & 0.008     &  5.36     & 36.64      &  0.18   &   122193  \\                
 $z =4$	             &  0.05    &       7.93              &  1.87      & 0.007     &  12.53    & 23.44      &  0.09   &   59092$^{1}$  \\                
 \hline
 \end{tabular}
 \tablebib{ (1) We note that Aq-D-SPH and Aq-E-SPH contain more than 100000 particles above $z=3$.}
 \end{table*}

\subsection{The evolution of mass distribution}\label{sec:massDistr}

\begin{figure}
\centering
 \includegraphics[width=0.45\textwidth]{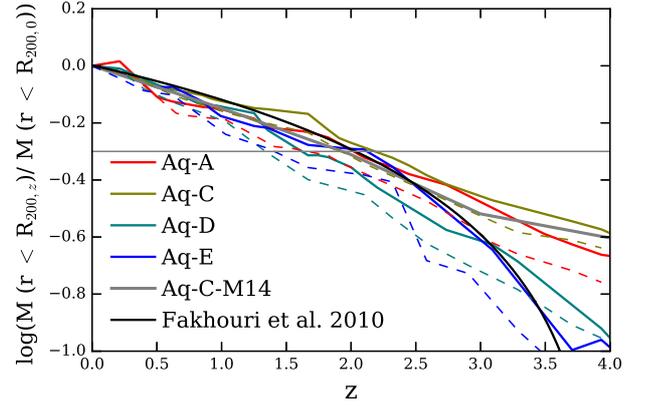}
  \caption{The mass accretion history of the haloes, computed as the mass of dark matter 
  enclosed within $r < R_{200,z}$ at redshift $z$, 
  normalized by the mass at $z=0$, as a function of redshift for SPH 
  (filled lines, Aq-A-SPH red, Aq-C-SPH brown, Aq-D-SPH green, and Aq-E-SPH blue) and DMO (dashed lines) counterparts.
  We also include the results from Aq-C-M14 (grey solid line). We compare our results with Eq.~2 from \citet{Fakhouri2010} (black line).
  Grey horizontal line is used as reference to estimate the formation time of the haloes, as the redshift at which the mass of the halo
  reach half of their mass at $z=0$. }
  \label{fig:MAH}
 \end{figure}

\begin{figure}
\centering
  \includegraphics[width=0.45\textwidth]{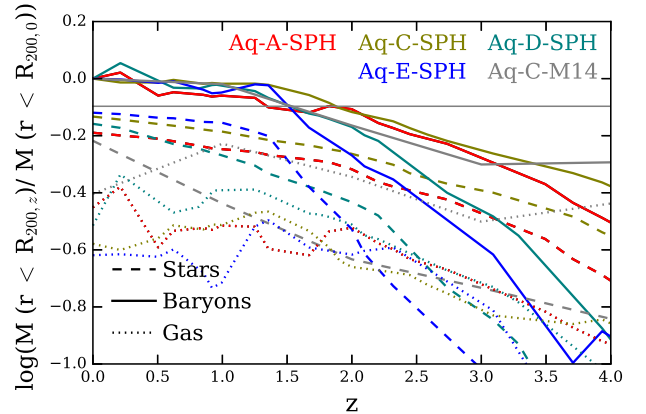}
  \caption{Mass accretion history of stars (dashed lines), and gas (dotted lines) enclosed within $r < R_{200,z}$, 
  normalized by the baryonic mass at $z=0$ as a function of redshift for the five haloes (Aq-A-SPH red,
  Aq-C-SPH brown, Aq-D-SPH green, Aq-E-SPH blue and Aq-C-M14 grey). We also plot the contribution of both together (i.e., baryons, filled lines). 
  The grey horizontal line is shown to give a reference of the redshift at which each halo reach the 80 percent of the present baryonic mass.}
  \label{fig:MAH-bar}
 \end{figure}

In order to gain insight into the influence of the assembly history on the halo properties, we compare the mass accretion history (MAH) 
of dark matter haloes within the virial radius in Fig.~\ref{fig:MAH}. We include the model proposed by \citet{Fakhouri2010} as a reference 
(see eq.~2 in the aforementioned work), who investigate the MAH from Millennium and Millennium-II simulations.
Our results show that the mass is accreted earlier in the SPH haloes. 

We define the formation time of each dark matter halo
as the redshift at which for the first time the halo reach half of its present mass (see the grey horizontal line in Fig.~\ref{fig:MAH}). 
Our findings show that SPH haloes are assembled earlier than their DMO counterparts.
This difference may be due to the impact of baryons. The SPH haloes got more 
concentrated earlier and in turn accrete more mass earlier than the DMO haloes.
Interestingly, haloes Aq-A-SPH and Aq-C-SPH assemble before, between $z \sim 2 - 2.3$, than the haloes Aq-D-SPH and Aq-E-SPH which
assemble at $z < 2$.
A similar trend for the mass assembly histories can also be seen in the DMO haloes, thus Aq-A-DMO and Aq-C-DMO
assemble before than Aq-D-DMO and Aq-E-DMO \citep[see also][]{Wang2011}.
We also find that the halo Aq-C-M14 assemble later than Aq-C-SPH,  
as a consequence of the regulation of the star formation activity (see ahead in this section).
The MAH of Aq-C-M14 compared with its DMO counterpart show only slight differences in time. 

To get further insight into this process, in Fig.~\ref{fig:MAH-bar} we analyse the mass accretion history of baryons from the SPH runs
for the stellar and gas components.
At $z=0$ these galaxies have baryonic mass relative to halo mass in the range of $\log(M_{\rm bar,0}/M_{200,0}) \sim -0.9$ to $-1.1$.
At $z \gtrsim 2.0$, the halos Aq-A-SPH and Aq-C-SPH present a larger fraction of stellar mass compared to Aq-D-SPH and Aq-E-SPH, while
at lower redshifts ($z \lesssim 2.0$) the trend flattens. This is more clear
for the total baryonic content, revealing that it remains roughly constant between $z\sim 0-2$.
Combining this result with those from Fig.~\ref{fig:DM-profiles2}, we find that there is a remarkable connection between the baryonic
mass content and the stability of the inner dark matter halo profile in time.

For instance, we find that at $z\sim2$ the $\sim$80\% of the present baryonic mass 
is assembled for Aq-A-SPH and Aq-C-SPH which in turn are the ones that
assemble earlier and get a steady dark matter density profile at this stage. 
These haloes reach a stellar mass which represent $\sim 60$\% of the total baryonic mass at present. 
The haloes Aq-D-SPH and Aq-E-SPH reach the 80\% of the present baryonic mass and
$\sim 50-60$\% of stellar mass with respect to the present baryonic mass at $z<2$.
Moreover, all the haloes show that $\log(M_{\rm bar,z'}/M_{200,0}) \sim -1.22 - -1.15$ at the redshift $z'$ they reach
the 80\% of the present baryonic mass.
This redshift is also related to
the time at which these haloes reach a stable dark matter density profile in the inner regions.
Our findings suggest that the baryons and in particular the 
stellar mass in these haloes, play a key role in the evolution of the dark
matter haloes. When the galaxy reaches a stellar mass compatible with $\sim 60$\% of the total baryonic mass at $z=0$,
the halo get a more stable and concentrated dark matter density profile.
These findings are in agreement with previous results showing that stellar mass is the main ingredient that affects 
the dark matter distribution within the haloes \citep{Duffy2010,Pedrosa2010,Tissera2010,Zemp2012,DiCintio2014b,Ceverino2015,Dutton2016,Tollet2016}.

The halo Aq-C-M14 shows a similar dependence with respect to the baryonic mass content and the evolution 
of the dark matter density profile.
We find that at $z \sim 1.5$,  $\sim$80\% of the present baryonic mass is assembled,
and the halo presents a steady dark matter density profile between $z=0-1$.
Furthermore, at $z\gtrsim1.0$ Aq-C-M14 shows a larger fraction of gas than stars.
This result is expected considering that the central
galaxy is more disc-dominated than the central galaxies in the SPH haloes. 
We note that at $z\sim1.5$ Aq-C-M14 reach $\sim 60$\% gas mass of the total baryonic mass at present. 
This result is connected with our findings for the SPH haloes and the stellar mass component of the galaxies.
Baryons appear to be fundamental for the evolution of the dark matter density profile
for haloes with $\log(M_{\rm bar,0}/M_{200,0}) \sim -0.9$ to $-1.1$ at z=0.

Another important ingredient that affects the mass distribution and evolution of the dark matter haloes is the merger history. 
\citet{Scannapieco2009} investigate the merger history of the SPH haloes between $z=0-3$ (see Fig. 9 of the aforementioned work)
and identify those satellites that enter either to the virial radius or the central 27~kpc.
Their results show that any of the SPH haloes studied in this work experience major mergers, 
and only intermediate or minor mergers occur in all the haloes\footnote{\citet{Scannapieco2009} 
define as intermediate mergers when the mass ratio is $M_{\rm sat}/M_{\rm cen} = 0.1-0.3$, 
and minor mergers $M_{\rm sat}/M_{\rm cen} = 0.02-0.1$, where
$M_{\rm sat}$ and $M_{\rm cen}$ are the stellar masses of the satellite and the central galaxy.}. 
In particular, Aq-D-SPH experiences a larger amount of minor mergers in time, while
from visual inspection we find that Aq-E-SPH experiences intermediate mergers between $z\sim 3$--4.
Furthermore, all the haloes endure minor and/or intermediate mergers within the 27~kpc between  $z=0-3$. 
Hence, our findings suggest that the differences in the formation times and distribution of the dark matter haloes might also be affected by their 
different merger histories. 

We also compare our findings with the stellar disc components of the central galaxies in each SPH halo presented in \citet{Scannapieco2009}, in order
to discuss other factors that might explain the evolution in the dark matter halo profiles. 
However, we do not find a dependency on this property. 
According to the definition adopted by \citet{Scannapieco2009}, all the haloes present a stellar disc component
which is more prominent for Aq-C-SPH, Aq-D-SPH and Aq-E-SPH.
We note that the halo Aq-C-M14 has a much more prominent disc component than Aq-C-SPH \citep{Marinacci2014}.

\begin{figure*}
 \centering
  \includegraphics[width=1.0\textwidth]{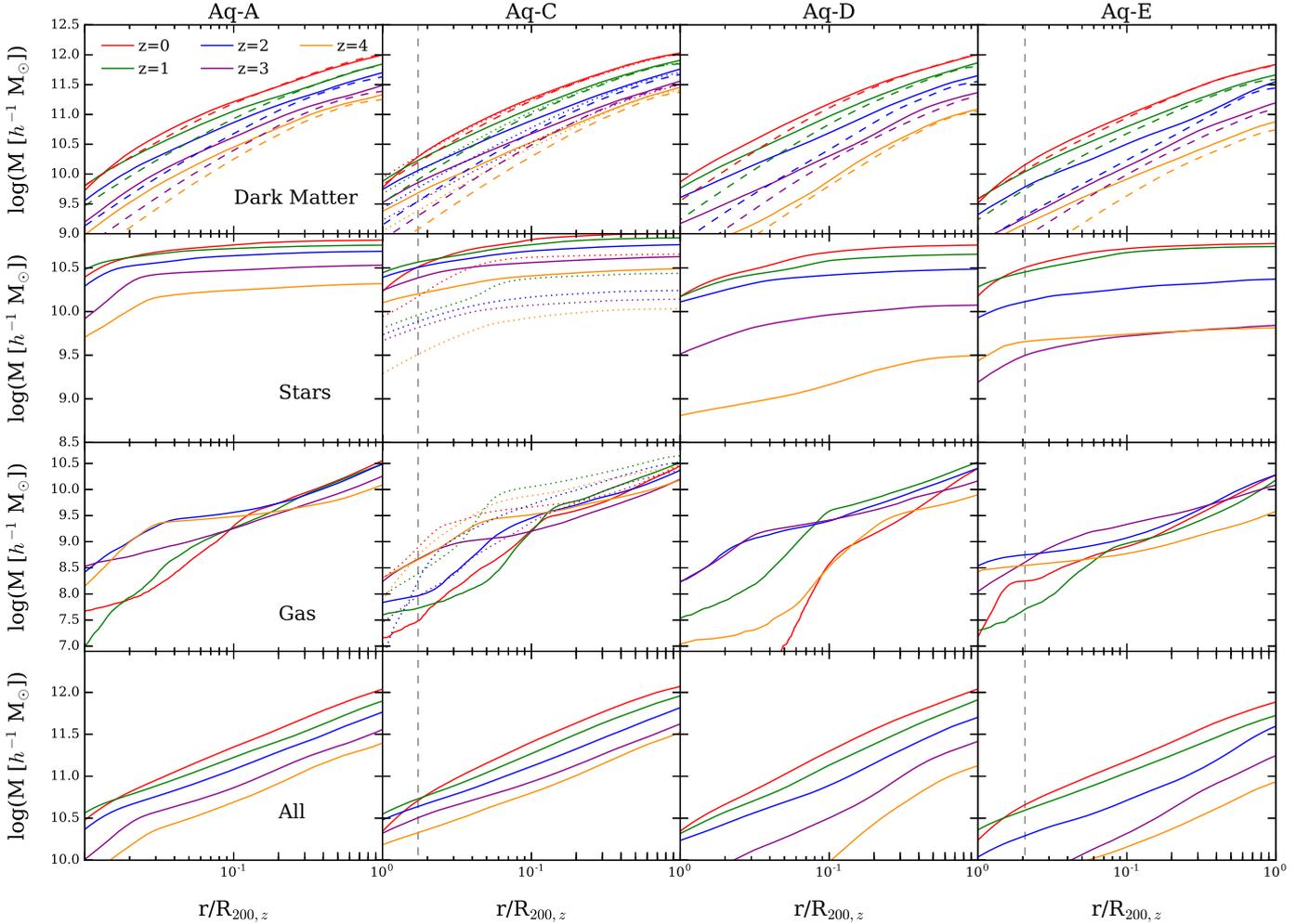}
   \caption{Cumulative mass distribution as a function of distance to the centre,
   and normalized by the virial radius of each redshift, R$_{\rm 200,z}$ for the four haloes studied in the hydro run 
   (filled lines). We also show the results from Aq-C-M14 (dotted lines in Aq-C panel).
   Grey dashed lines represent three times the gravitational softening in Aq-C-SPH and Aq-E-SPH at $z=0$,
   while in Aq-A-SPH and Aq-D-SPH this value is below the range we show in this figure.}
  \label{fig:cumularive-MD}
 \end{figure*}

In Fig.~\ref{fig:cumularive-MD} we present the cumulative mass distribution of the different components (stars, gas, dark matter and all)
as a function of distance at $z=0, 1, 2, 3$ and 4 for the SPH haloes and Aq-C-M14. The galactocentric distance
is normalized by the virial radius at each redshift R$_{200, z}$
to compare the mass distribution in time. We also include the cumulative mass from DMO haloes.
We find that dark matter concentrates and increases in the inner region of the haloes as redshift decrease for SPH and DMO haloes, 
as we stated previously in Fig.~\ref{fig:MAH}. Furthermore, the dark matter mass in the inner region is higher for the SPH haloes than for their DMO counterparts. This difference is also present in Aq-C-M14.

For the SPH runs, we find that each halo present different trends for the stellar and gas distribution, 
but in agreement with our findings from Fig.~\ref{fig:MAH-bar}. 
We find that all the SPH haloes and also Aq-C-M14, have higher stellar mass content within 5\%R$_{{\rm 200}, z}$ 
compared to the gas content.

As we stated previously, the evolution of the stellar content within the inner parts of the halo might play a role in  
the dark matter density profiles. 
Furthermore, Aq-A-SPH and Aq-C-SPH show that the stellar mass in the inner part of the haloes (within 5\%R$_{{\rm 200},z}$) is higher than 
the dark matter mass in the same region, while Aq-D-SPH and Aq-E-SPH present opposite relation. This finding reinforces
our assumption on the role that the stellar component has in the inner regions of the haloes.
Hence, Aq-A-SPH and Aq-C-SPH assembled earlier and concentrate a significant amount of stellar mass in the inner regions which in turn
determines the inner profile of the dark matter haloes, in agreement with previous results \citep{Duffy2010,Zemp2012}.

The cumulative distribution of gas in Fig.~\ref{fig:cumularive-MD} do not show a definite trend for the SPH haloes. 
Three of them have more gas in the inner region at $z=2$ and moving outwards at low redshift. 
These variations in the gas mass can be explained due to the contribution of star formation events together with outflows produced
by supernovae feedback. 
Also the gas accreted during merger processes affecting each halo can be relevant.

\subsection{Dark matter haloes shape}

 \begin{figure*}
 \centering
 \includegraphics[width=0.45\textwidth]{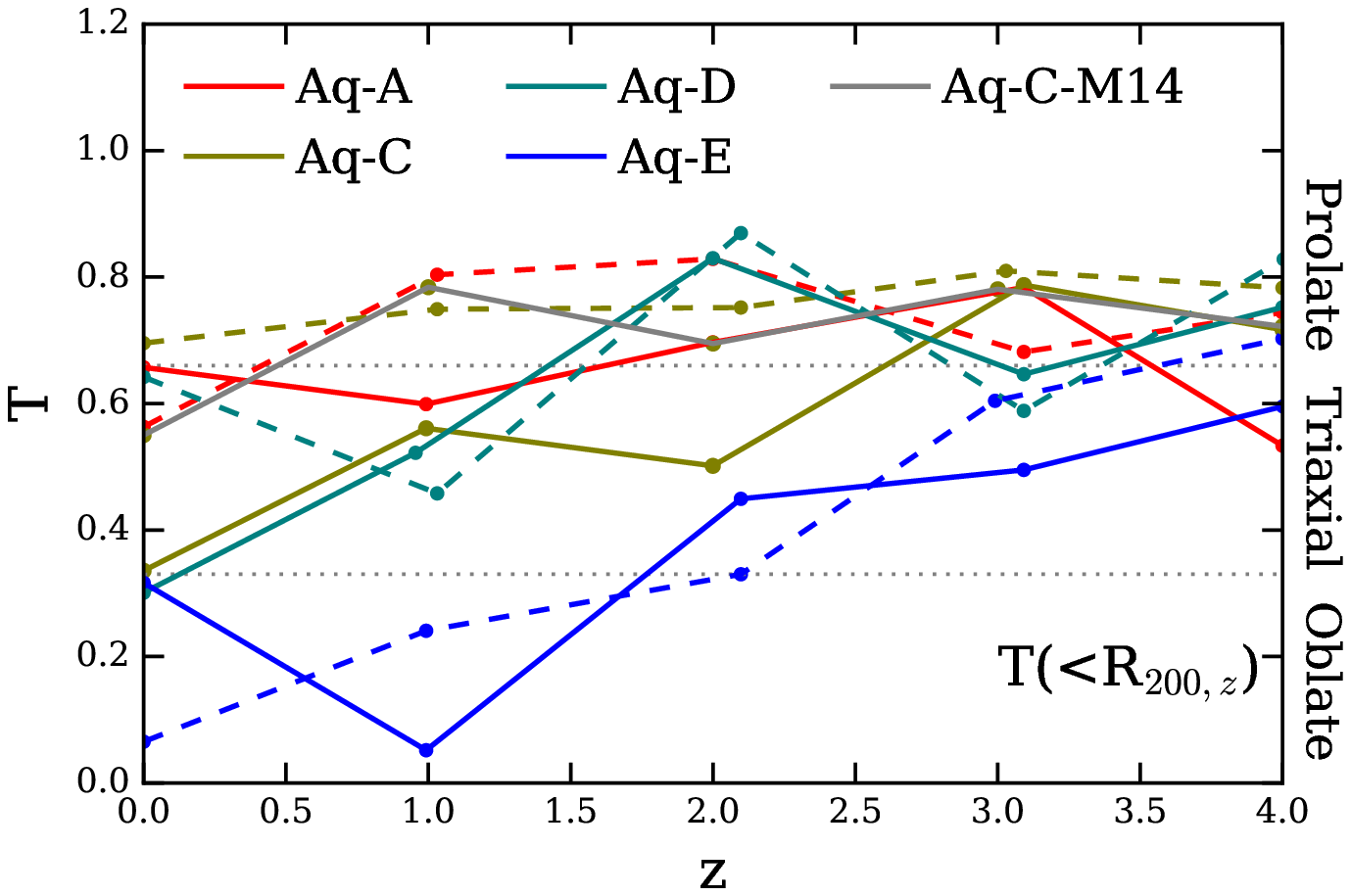}
 \includegraphics[width=0.45\textwidth]{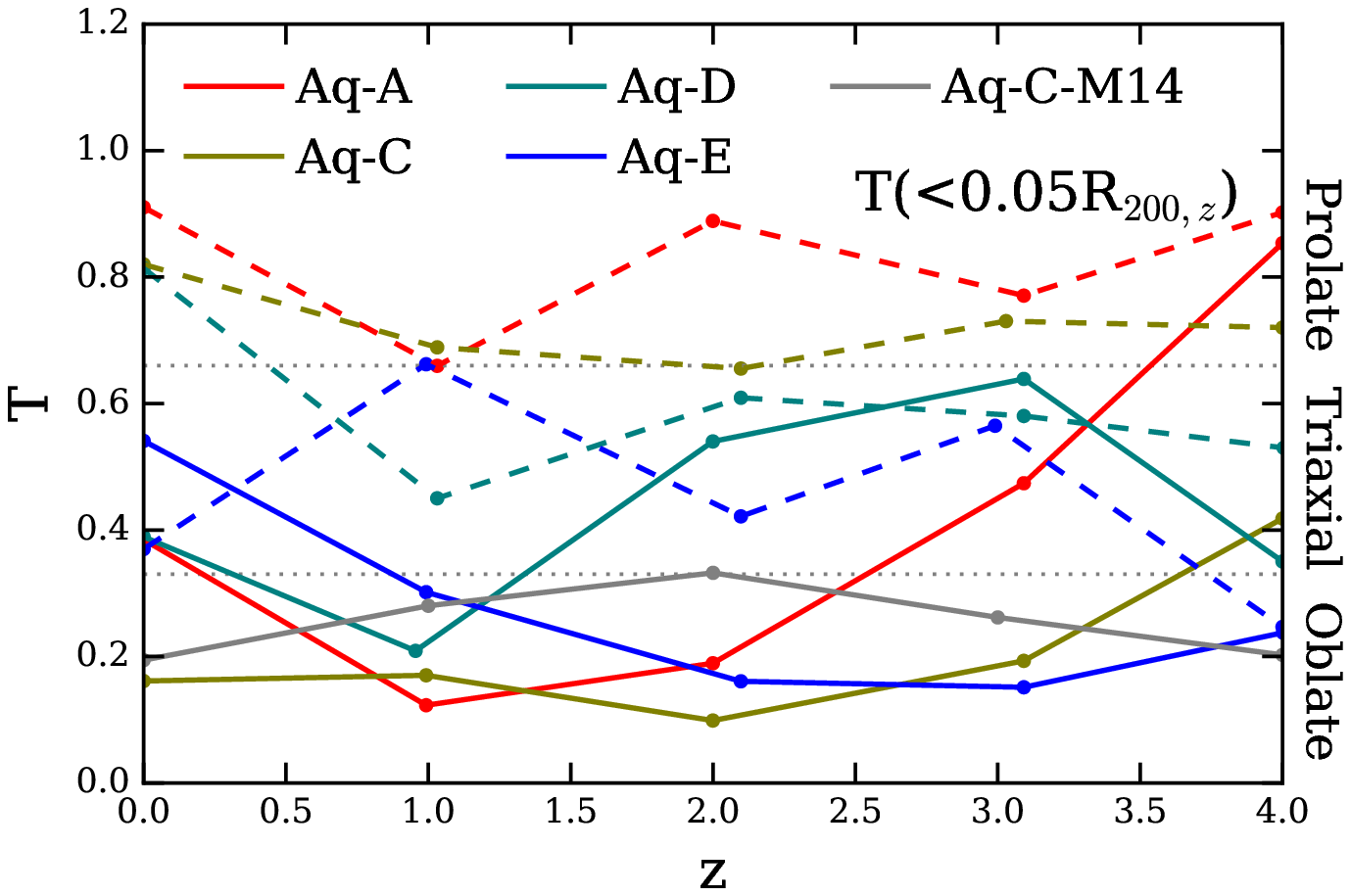}
   \caption{ Comparison of the triaxiality of the haloes as a function of redshift for the SPH (filled lines) 
   and DMO (dashed lines) haloes within the virial radius at each redshift R$_{200,z}$ (\textit{left panel})
   and 5\% of the virial radius, 0.05~R$_{200,z}$ (\textit{right panel}).
   We also include the results obtained from Aq-C-M14 (solid grey lines). Dotted grey lines represent the adopted 
   criterion to define oblate ($0< {\rm T}<1/3$), triaxial ($1/3 < {\rm T} < 2/3$) and prolate ($2/3 < {\rm T} < 1$) shapes.}
   \label{fig:TriaxRvir}
 \end{figure*}  
 
N-body simulations have shown that within a CDM scenario, dark matter haloes are typically triaxial
\citep[e.g.,][]{Frenk1988,Jing2002,Allgood2006}, with prolate shapes in the inner regions and more oblate in the outskirts. 
Their original morphology would be due to the haloes assemble along a preferred 
direction dominated by the cosmic-web environment \citep{Avila-Reese2005,Patiri2006,Vera-Ciro2011}.
However, as baryons condense within the inner regions of the dark matter haloes, they dominate the potential at the centre and may 
affect the DM-haloes shape making them more oblate systems \citep{Tissera1998,Kazantzidis2004,Tissera2010,Butsky2016}.

We investigate the shape evolution of the dark matter haloes from the SPH and DMO runs, employing the same 
methodology than in \citet{Tissera2010}, based on the one proposed in \citet{Dubinski1991}.
Hence, we compute the eigenvalues of the tensor of inertia and obtain the semi-axes of the triaxial ellipsoids (a$>$b$>$c).
The triaxiality of the dark matter haloes is defined as T$=\frac{a^{2} - b^{2}}{a^{2} - c^{2}}$. 
We adopt the following definition: the shape is oblate if
$0<{\rm T}<1/3$, prolate if $2/3 < {\rm T} < 1$, and triaxial if $1/3 < {\rm T} < 2/3$.
The morphology and the evolution with redshift of the DMO haloes of Aquarius project is also studied by \citet{Vera-Ciro2011} 
and \citet{Zhu2017} using different methodologies.

We analyse the evolution of the triaxiality within the virial radius at each redshift R$_{200,z}$ and
within 5\%R$_{200,z}$, in order to investigate the inner and global shape of the haloes.  
We show our results at $z = 0, 1, 2, 3$ and 4 for the SPH haloes and their DMO counterparts. 

On the left panel of Fig.~\ref{fig:TriaxRvir}, we present the triaxiality within R$_{200,z}$. 
In agreement with \citet{Vera-Ciro2011}, the shape of DMO haloes show to be prolate at high redshifts and more triaxial at low redshifts.
We find that Aq-E-DMO presents a much remarkable change with redshift, compared with the rest of DMO haloes, 
reaching more oblate shape at $z=0$.
The differences between the DMO shapes and their evolution is discussed in \citet{Vera-Ciro2011} who
claims that at high redshift the haloes are mainly influenced 
by the mass assembly and environment, with accretion of 
matter through filaments. At low redshift, the accretion is isotropic and the haloes evolve to oblate configuration.
The mass accretion in Aq-E-DMO is more isotropic during cosmic time, which would be the reason  
this halo presents a more oblate shape than the others.

For the SPH haloes, we find that Aq-C-SPH, Aq-D-SPH and Aq-E-SPH are prolate at high redshift, becoming more oblate at low redshifts.
This trend is slightly stronger than in the DMO counterparts for Aq-C-SPH and Aq-D-SPH. 
Aq-E-SPH is more triaxial than Aq-E-DMO, showing that baryons might modify the global shape in a way opposite than expected.
Aq-A-SPH tend to be in general more prolate than the others. 

From the comparison between Aq-C-SPH and Aq-C-M14, we find that the later is more prolate in the redshift range studied.
It is interesting to note that the two haloes that contain the most prominent stellar disc
(Aq-A-SPH and Aq-C-M14) have also a more prolate shape in time. 
Interestingly as the central galaxy is confined to the very central region of the halo, 
their effect into the global shape of the halo should be negligible. 

On the right panel of Fig.~\ref{fig:TriaxRvir} we show the triaxiality in the inner regions of the haloes, within 5\%R$_{200,z}$.
We find that the SPH haloes are more oblate, including also Aq-C-M14, while the DMO counterparts are more prolate in all the redshifts studied.
This result is expected and in agreement with previous findings \citep{Tissera2010,Zemp2012,Ceverino2015,Butsky2016}, 
due to the impact of baryons in the central regions. 
As stellar mass condense in inner parts of the SPH haloes, the shape of the dark matter haloes
tend to be more rounder and oblate. 
Previous works have investigated how the orbital properties of dark matter haloes change due to the 
baryon contraction in the central regions \citep{Debattista2008,Zhu2017}.
In particular, \citet{Zhu2017} use the halo Aq-C to explore this aspect,
finding different behaviour for the dark matter particle orbits when baryons are included.  
The presence of baryons in the inner regions as galaxy collapse to the centre,
modify the dark matter trajectory from box orbits into tube or rounder orbits turning the inner shape of the halo more oblate.

We also compare our findings considering the differences and similarities of the dark matter halo profiles.
Our results show that Aq-A-SPH and Aq-C-SPH present only partial similarities.
These haloes show that the triaxiality reduces between $z=4$ and $z=1$ becoming more oblate, 
while at $z=0$ the trend reverts for Aq-A-SPH turning to triaxial. 
We find that Aq-C-M14 is more oblate in all the redshift range studied.
Although Aq-C-M14 contains a much more prominent disc than Aq-C-SPH, this does not seem to affect the inner shape.

\subsection{Angular momentum of dark matter}

  \begin{figure}
 \centering
 \includegraphics[width=0.4\textwidth]{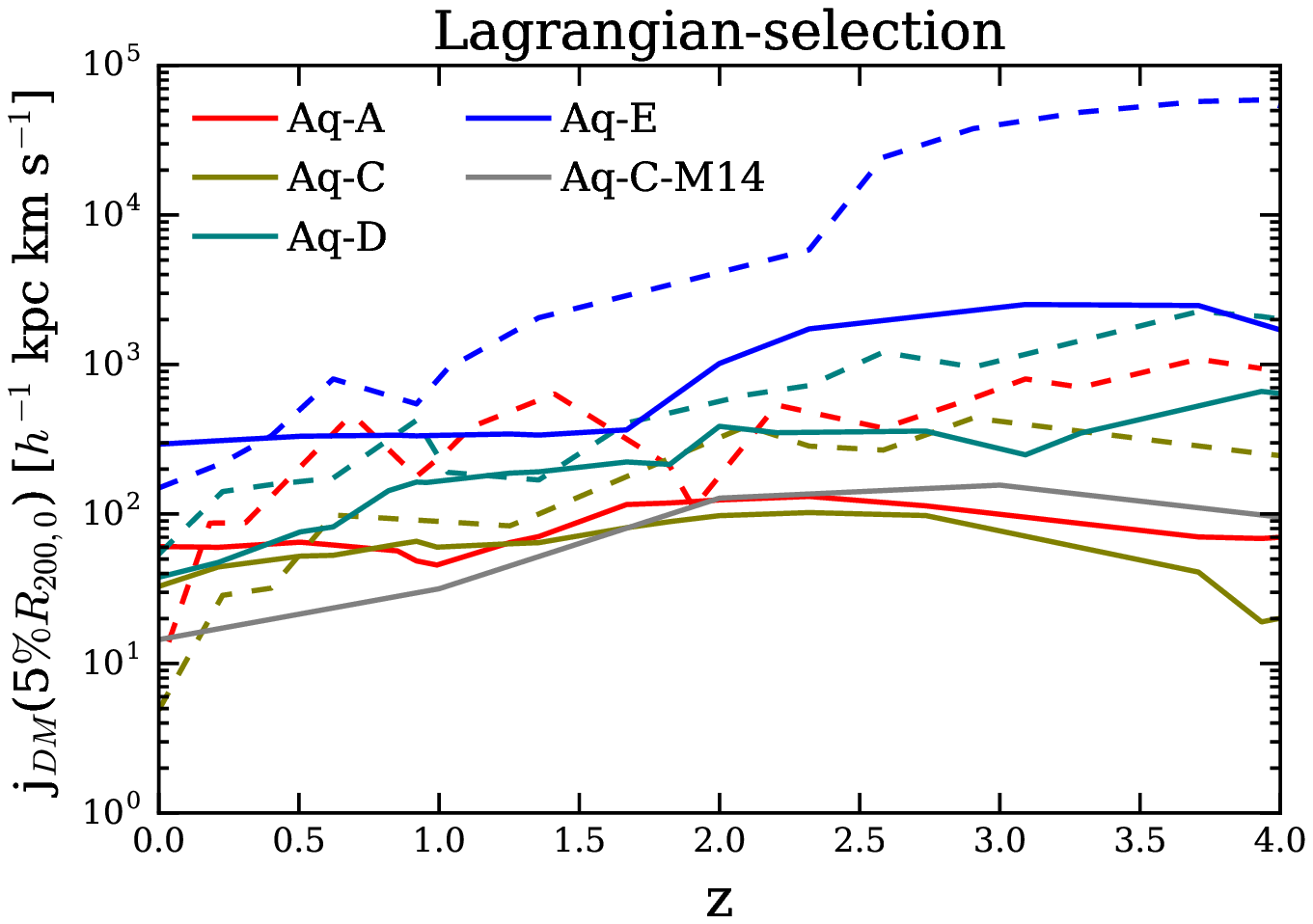}   
 \includegraphics[width=0.4\textwidth]{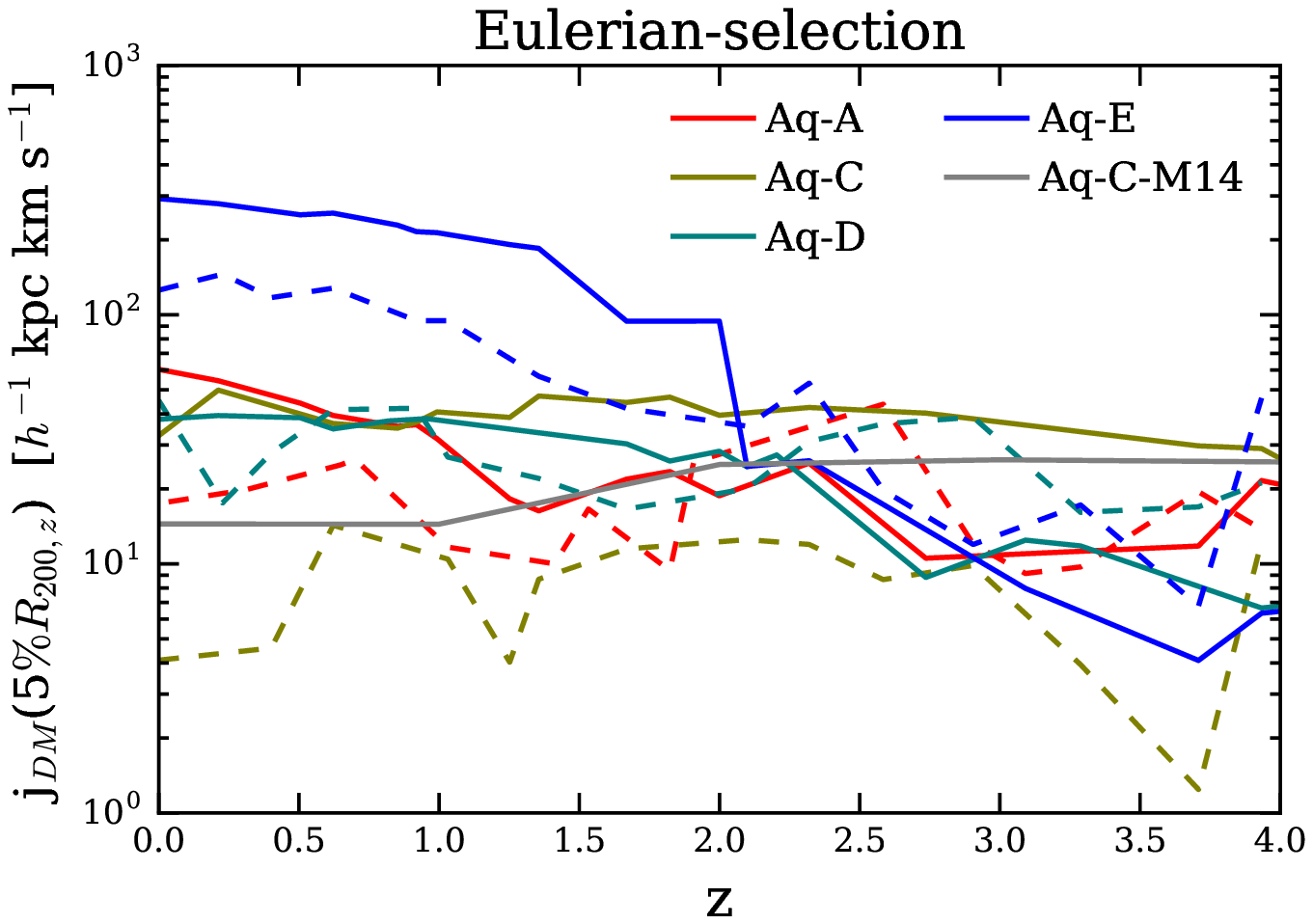} 
 \caption{Specific angular momentum of dark matter particles for the haloes studied. 
 \textit{Top panel:} we compute it through Lagrangian-selection within the 5\%R$_{\rm 200}$ at $z = 0$. Therefore, 
 we follow the same particles and compute the specific angular momentum for each redshift.
 We present our results for the SPH haloes (filled lines) and their DMO counterparts (dashed lines). We also include
 the halo Aq-C-M14 (solid grey line).
 \textit{Bottom panel:} we compute it through Eulerian-selection within the 5\%R$_{\rm 200,z}$ as 
 a function of redshift for the SPH haloes (filled lines) and their DMO counterparts (dashed lines). We also include
 the halo Aq-C-M14 (solid grey line).}
 \label{fig:AngMom-DM}
 \end{figure}

  \begin{figure}
  \centering
  \includegraphics[width=0.4\textwidth]{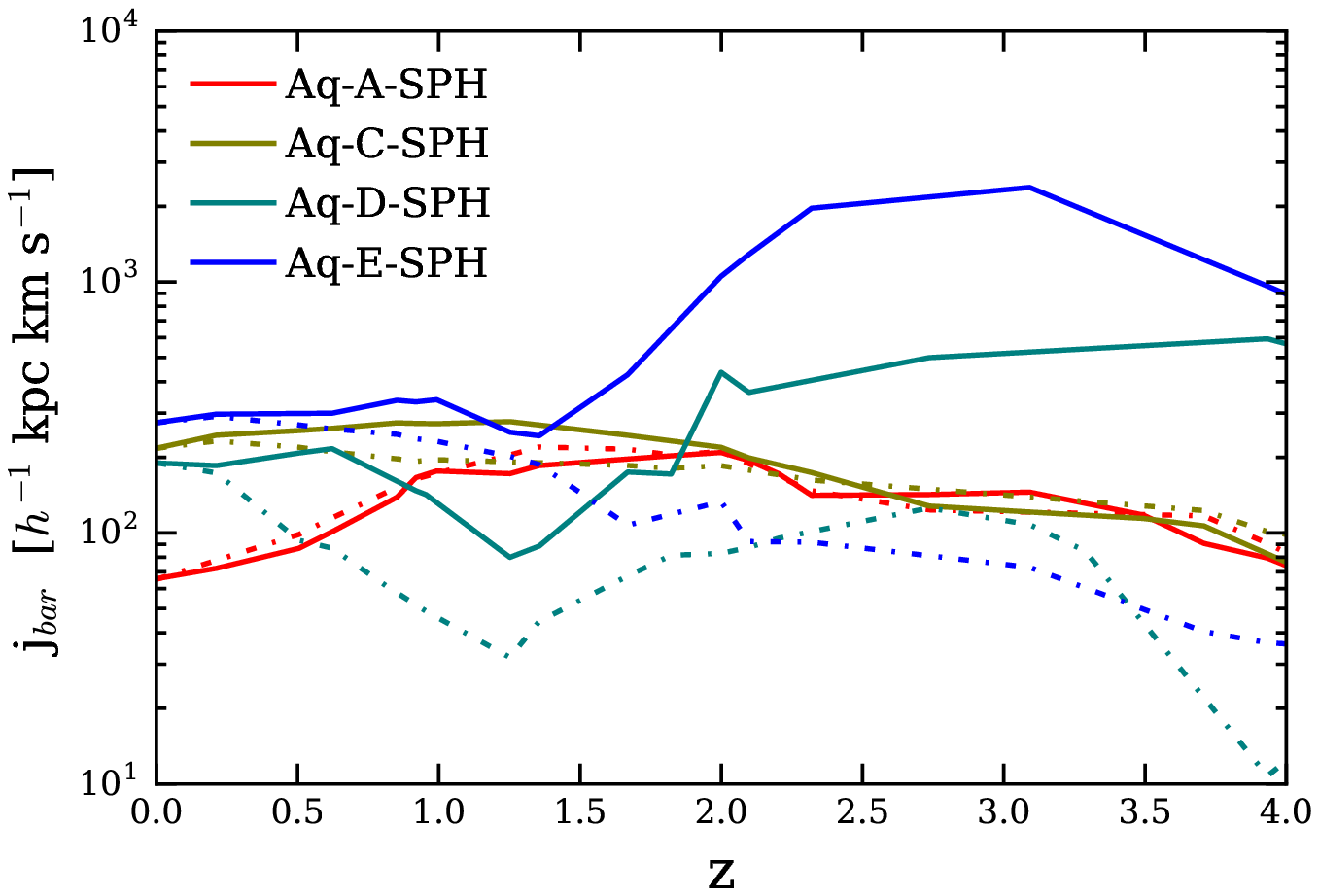}   
  \caption{Specific angular momentum of baryons as a function of redshift
  through Lagrangian (filled lines) and Eulerian (dotted dashed lines) selection, using the same methods than for the
  dark matter particles.}
  \label{fig:AngMom-bar}
  \end{figure}

The evolution of specific angular momentum of the dark matter haloes is also influenced 
by the baryon collapse in the central galaxy \citep{Pedrosa2010}. 
We define the specific angular momentum as 
$ \vec{j} = \frac{\sum_i m_i \vec{r_i} \times  \vec{v_i}}{\sum_i m_i}$,
where $r_i$ is the radial distance of the particle $i$ to the centre of mass of the halo,
$v_i$ is their peculiar velocity and $m_i$ is the mass of the particle $i$.

The specific angular momentum can be measured through Lagrangian-selection or Eulerian-selection.
In Lagrangian-selection, we identify the particles within a selected region at $z=0$ and track them back in time.
For the Eulerian-selection, we select a region in space or radial bin of the halo
and explore how the specific angular momentum evolves as a function of time.
In order to contrast the differences in the inner region of the haloes, we explore the specific angular momentum
using a Lagrangian-selection that track the dark matter particles within 5\%R$_{200,0}$ (at $z=0$) in time, and using
an Eulerian-selection within  5\%R$_{200,z}$ (at redshift $z$). 

We present our findings for the dark matter particles from the SPH and DMO runs in Fig.~\ref{fig:AngMom-DM},
using the Lagrangian-selection (\textit{top panel}) and Eulerian-selection (\textit{bottom panel}).
For all the DMO haloes, the Lagrangian-selection shows that the specific angular momentum decreases as
redshift decrease, reducing it around $\sim 85 - 95$~\% 
between $z=0-4$. This trend indicates that the dark matter halo particles loss angular momentum in the inner regions, 
as haloes concentrate in agreement with our findings from the dark matter halo profiles. 
In the case of the SPH haloes, we find two different behaviours in the Lagrangian-selection. The specific angular momentum of 
Aq-A-SPH and Aq-C-SPH is approximately constant, while for Aq-D-SPH and Aq-E-SPH decreases with redshift.
The differences can be explained since Aq-A-SPH and Aq-C-SPH assemble earlier and the stellar mass within 
the inner regions is large enough to stabilize the halo. 
To get more insight about this trend, we also compute the percentage of particles from the Lagrangian-selection that are
within 5\%R$_{200,0}$ as a function of redshift.   We find that for Aq-A-SPH and Aq-C-SPH 
 around $\sim 70-80\%$ of the dark matter particles tracked in the Lagrangian-selection are already within 
 5\%R$_{200,0}$ at higher redshifts. This explains why the specific angular momentum of these haloes is preserved in time. 
 For the halo Aq-D-SPH we find that the percentage of particles within the  5\%R$_{200,0}$ increase as redshift decrease, as halo concentrates to the inner region of the halo. 
 The halo Aq-E-SPH show also a similar trend than Aq-D-SPH between $z\sim 1.5-4$ mainly due to the impact of an intermediate merger that occur between $z=3-4$.
 For this halo, at redshifts $z < 1.5$ we find that the specific angular momentum is preserved, and 
 the number of dark matter particles from the Lagrangian-selection that are within the 5\%R$_{200,0}$ is roughly $\sim 70$\%.
 These results are in agreement with the evolution of dark matter halo profiles and the cumulative mass distribution presented in
 Fig.~\ref{fig:DM-profiles2} and Fig.~\ref{fig:cumularive-MD}, respectively.
In general terms for the Lagrangian-selection, we find that the DMO haloes present a higher specific angular momentum  
than their SPH counterparts. These differences might be explained since the SPH haloes assembled earlier than the DMO haloes, as a consequence of their
interaction with the baryonic matter in the central region of the SPH haloes \citep{Pedrosa2010}.
We also show the specific angular momentum for the halo Aq-C-M14. For the Lagrangian-selection, our results
present slight variations but a roughly constant specific angular momentum.

Recently, \citet{Zavala2016} investigate the evolution of the specific angular momentum and their connection with 
the morphology of the galaxies, following a Lagrangian-selection of particles within 10\% the virial radius at $z=0$. 
Their results indicate that after turnaround ($z\sim 4$), the inner 
dark matter haloes lose 90\% of its specific angular momentum. 
This reduction might be due to the transference of angular momentum to the outskirts of the halo during its assembly history, 
dominated by substructures merging onto the halo \citep[see also,][]{Frenk1985,Zavala2008}. 
These results are in agreement with our findings for Aq-D-SPH and Aq-E-SPH which in turn are the haloes that have a much more significant merger history. 

The Eulerian-selection shows that the specific angular momentum slightly increases as redshift decreases for 
Aq-A-SPH, Aq-C-SPH and Aq-D-SPH, while the halo Aq-E-SPH show a more significant change possibly due to the intermediate
mergers it experiences between $z\sim 3-4$. 
The halo Aq-C-M14 show a similar trend than the Lagrangian-selection method.
For the DMO counterparts, we do not find a clear trend for the specific angular momentum.
Hence, the results from the Eulerian-selection method are much difficult to interpret since
different number of dark matter particles are mapped as a function of redshift.

In Fig.~\ref{fig:AngMom-bar} we show the specific angular momentum for the baryons following the two 
same methods implemented for dark matter particles.
From the Lagrangian-selection, we find that the baryons within Aq-D-SPH and Aq-E-SPH lose $\sim 70$\% of their specific angular momentum between $z=0-4$,
while Aq-A-SPH and Aq-C-SPH do not present significant variations.
\citet{Zavala2016} discuss the differences of the specific angular momentum of baryons according to the morphology of the galaxy, finding that 
bulge-dominated galaxies loss in average 80\% of its angular momentum, while disc-dominated galaxies loss 50\%.
According to the definition adopted by \citet{Zavala2016}, the four central galaxies in the SPH haloes are bulge-dominated
(galaxies with $B_{*}/T_{*} > 0.5$ are bulge-dominated, where
$B_{*}$ is total stellar mass in the bulge, and $T_{*}$ is the total stellar mass; otherwise, the galaxies are disc-dominated). The results obtained from
Aq-D-SPH and Aq-E-SPH are in agreement with these findings considering the errors, while Aq-A-SPH and Aq-C-SPH do not follow this trend.
The Eulerian-selection present a slight increase of the specific angular momentum for Aq-D-SPH and Aq-E-SPH, 
showing that baryons collapse to the centre in agreement with the results from Lagrangian-selection. The haloes Aq-A-SPH and Aq-C-SPH show slight variations, and similar results between Eulerian and Lagrangian selection.

\subsection{The role of outflows and inflows}
\label{sec:ac}
 \begin{figure}
 \centering
    \includegraphics[width=0.4\textwidth]{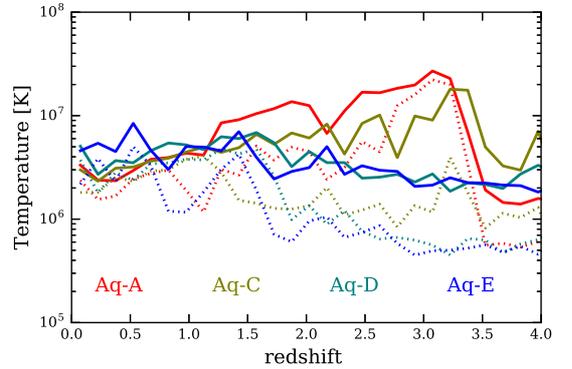}
   \caption{Mean temperature of the gas particles \textit{promoted} to the hot phase as a function of redshift 
   \citep[for details regarding this criteria see][]{Scannapieco2006}.
   Dotted lines represent the mean temperature in each redshift, while filled lines
   belong to the mean temperature considering only those gas particles with temperature above $T>10^6~K$. }
   \label{fig:promoted}
 \end{figure}  
 
   \begin{figure*}
 \centering
    \includegraphics[width=0.45\textwidth]{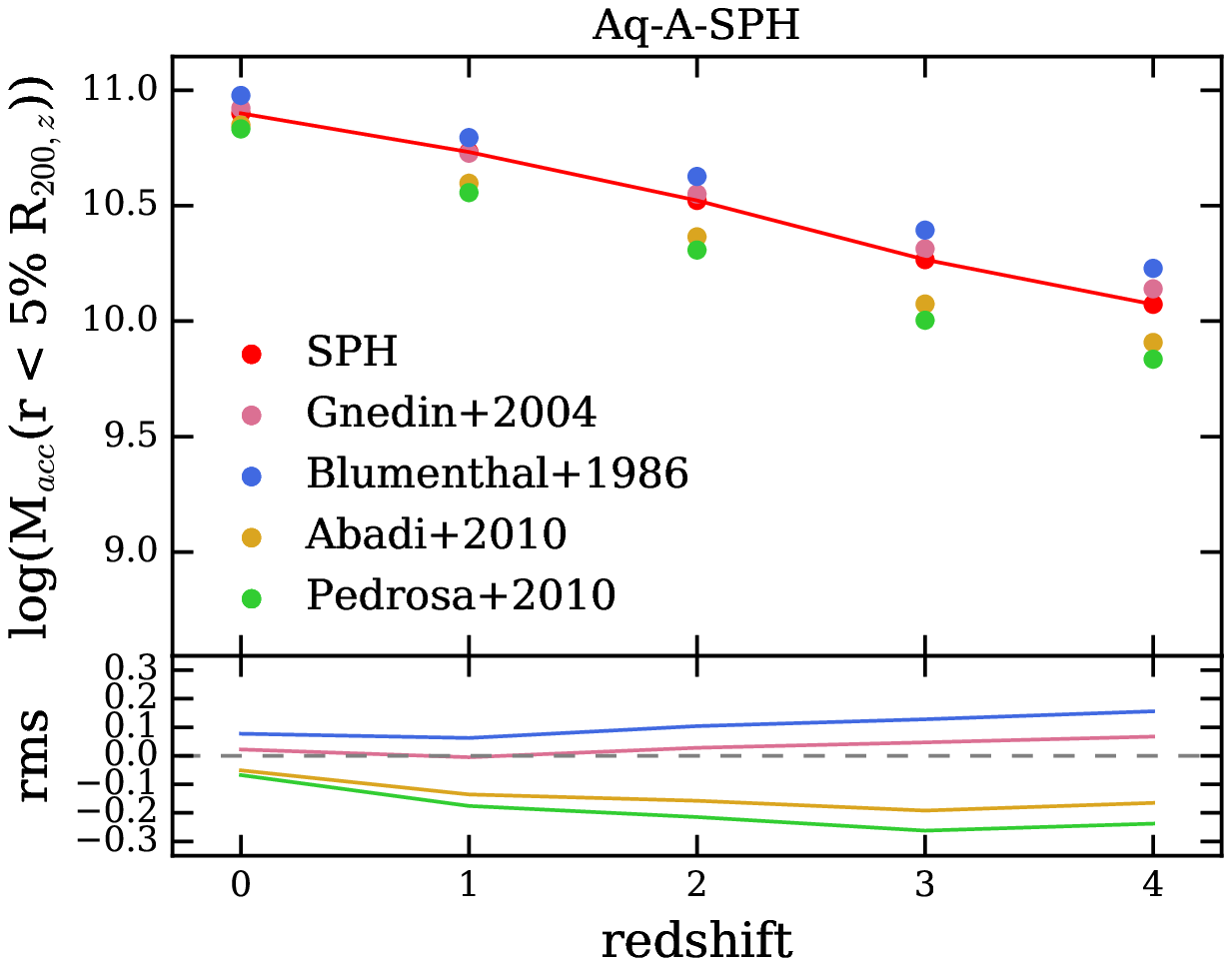}
    \includegraphics[width=0.45\textwidth]{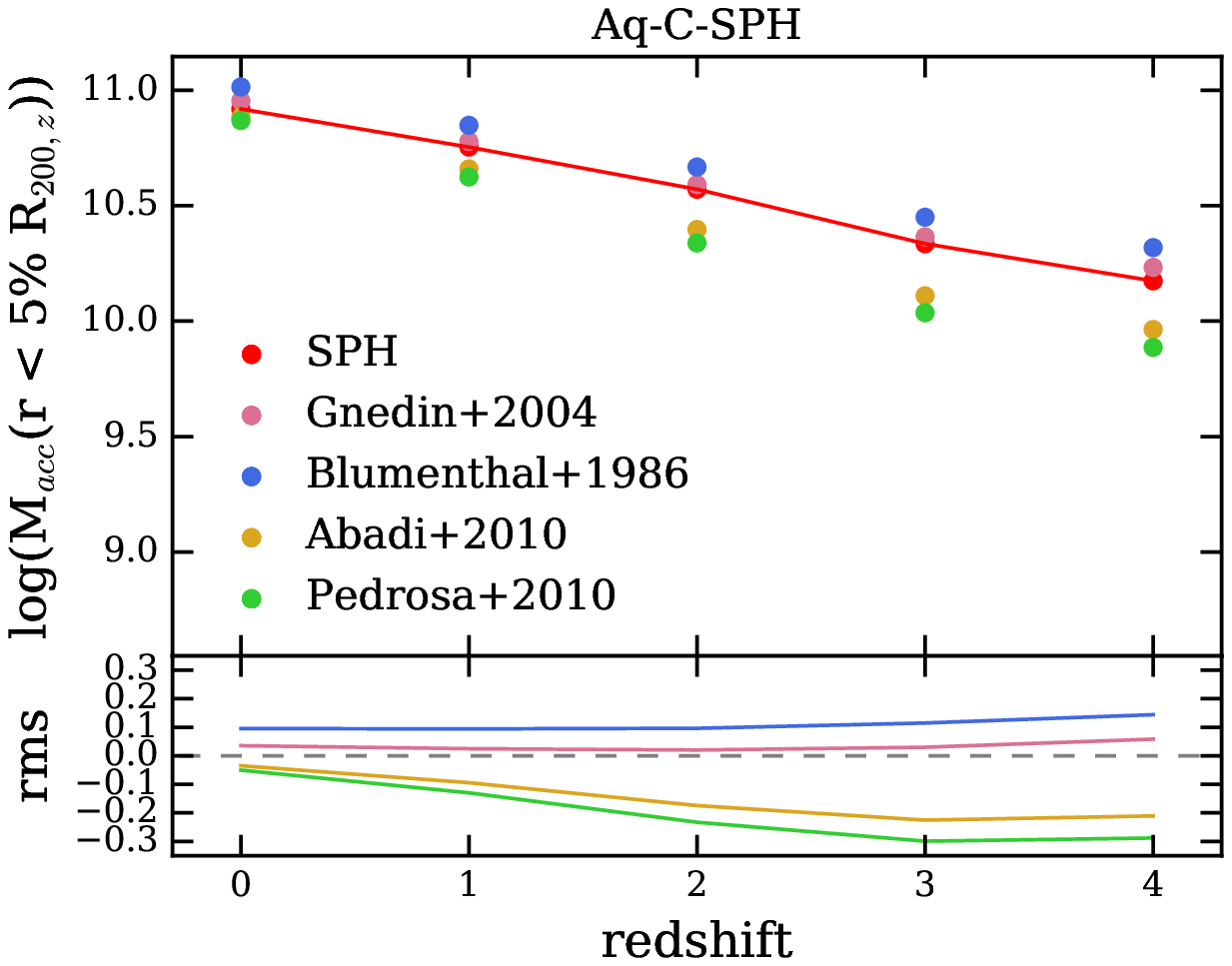}
    \includegraphics[width=0.45\textwidth]{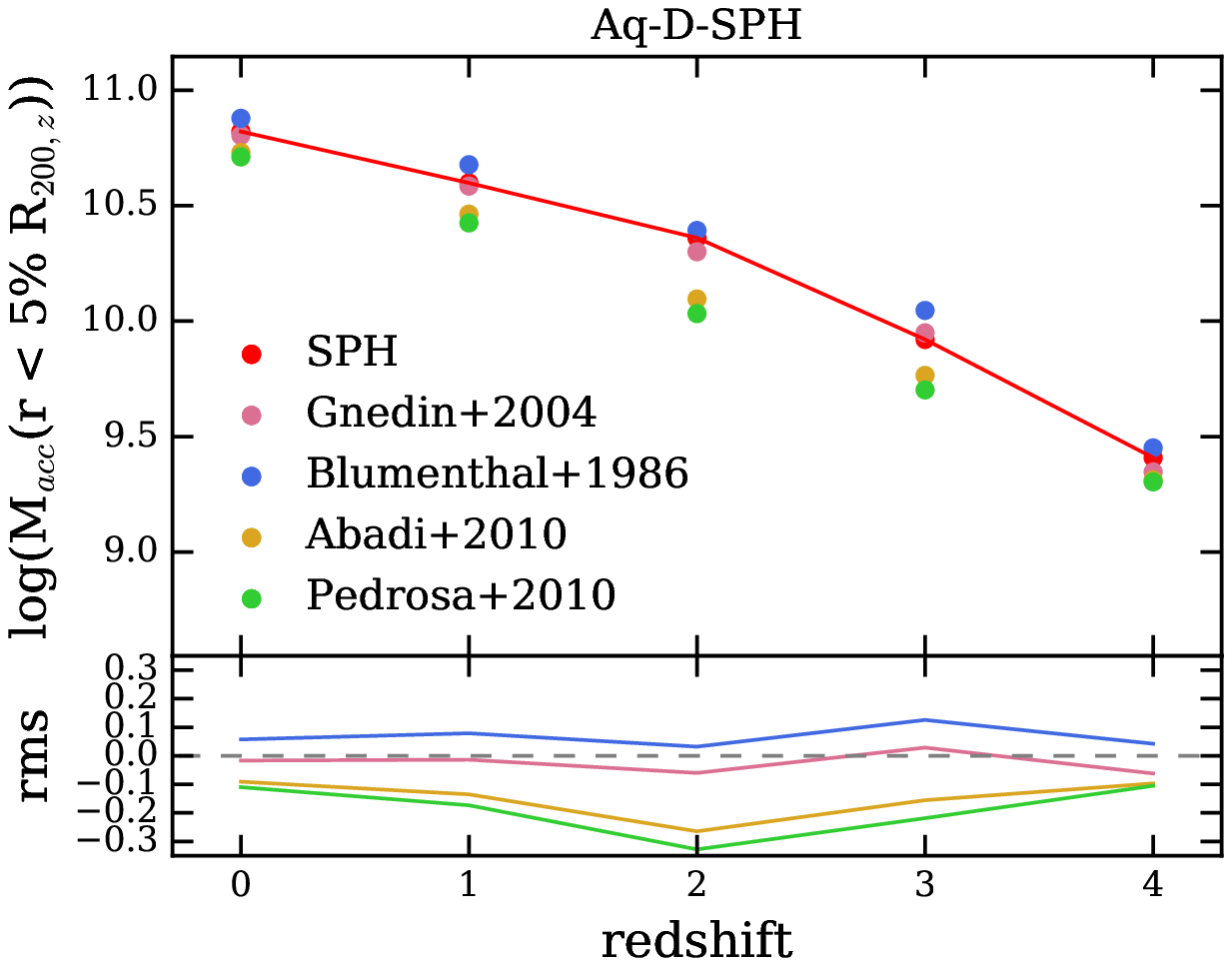}
    \includegraphics[width=0.45\textwidth]{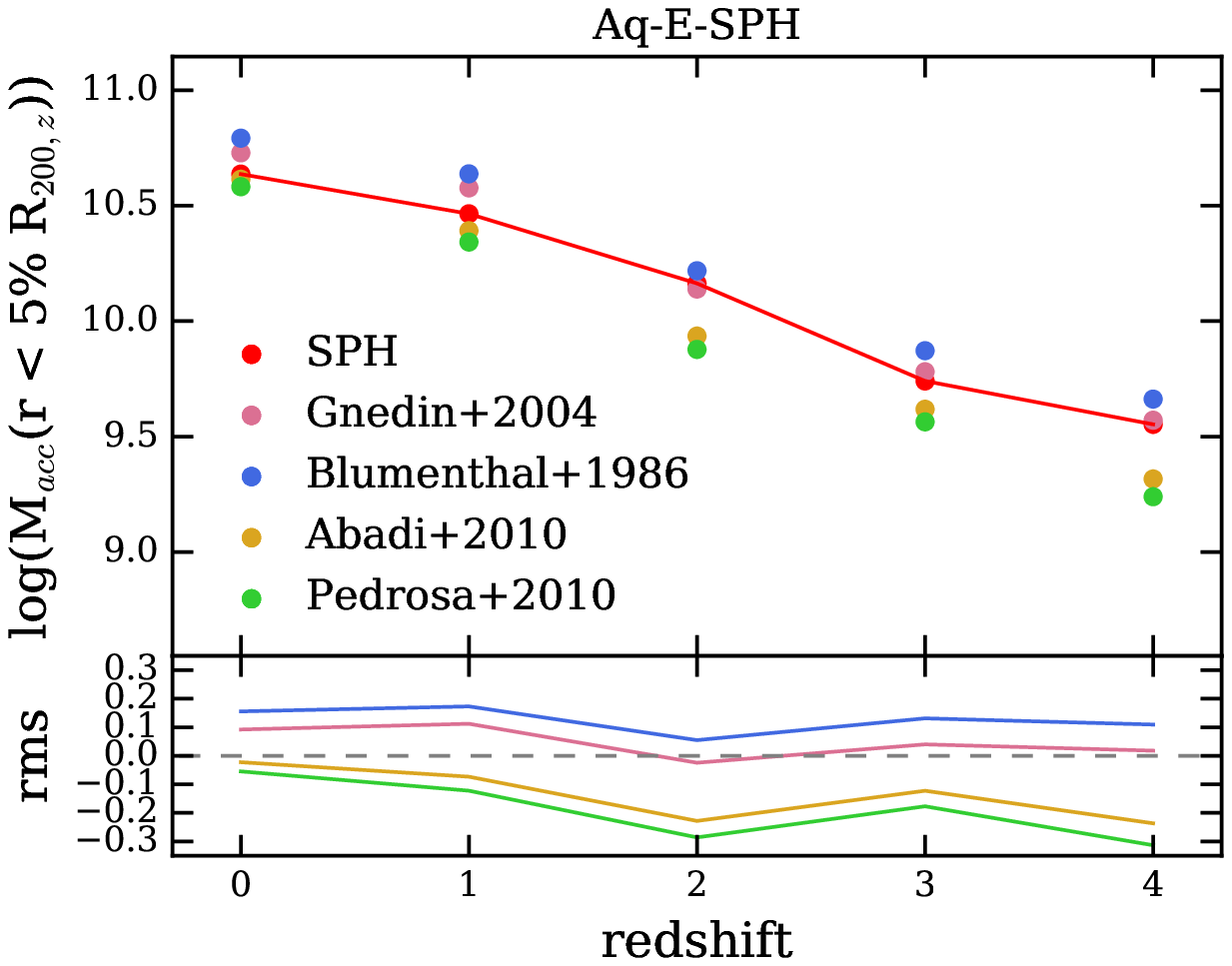}
   \caption{Mass enclosed within the five percent of the virial radius (5\%R$_{200,z}$) at redshifts $z=0$, 1, 2, 3 and 4.
   We show the results obtained from each halo (red points and lines) and those inferred from the AC implementations of
  \citet{Blumenthal1986} (blue points), and \citet{Gnedin2004} (pink points), and empirical models of \citet{Abadi2010} 
  (orange points) and \citet{Pedrosa2010} (green points).
  We also present the rms obtained from the comparison between the models and the SPH haloes.}
   \label{fig:AC}
 \end{figure*} 
 
    \begin{figure}
 \centering
    \includegraphics[width=0.4\textwidth]{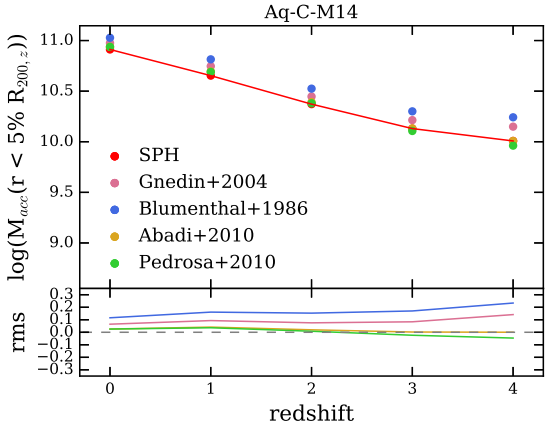}
   \caption{Same than Fig.~\ref{fig:AC} for Aq-C-M14.}
   \label{fig:AC-M14}
 \end{figure}

Mixed processes behind the galaxy formation and evolution such as gas inflow and outflows
can impact the inner region of the dark matter haloes \citep{Pedrosa2009,Pedrosa2010,Governato2012,Pontzen2014,Dutton2016,Tollet2016,Zhu2016}.
Furthermore, different reports
show that their impact depends on the mass of the dark matter haloes and the star formation rate of the galaxies within them.
For instance, dwarf galaxies present cores due to starbursts and subsequent supernovae feedback may harshly impact in the inner regions
producing an irreversible effect \citep{Pontzen2014}. On the other hand, many reports suggest that there is no core 
in Milky-Way type haloes, mainly due to the fact that the outflows may not be strong enough to
modify the potential wells of such dark matter haloes. Nonetheless, high-resolution simulations
are needed to study this feature in detail \citep{Maccio2012,Mollitor2015}.

In this section we discuss the impact of baryons on the dark matter haloes, by analysing two aspects: the galactic outflows
produced by supernovae feedback, and the baryonic contraction within the SPH haloes.
The cosmic star formation rate of the SPH runs implemented in this work are analysed in \citet{Scannapieco2009} (see Fig. 5).
\citet{Scannapieco2009} show that Aq-A-SPH and Aq-C-SPH have a peak of star formation before $z\sim 3$ 
while Aq-D-SPH and Aq-E-SPH show this peak at a later time.
The star-forming bursts produce an increase of the SNe feedback sources which in turn modify the temperature of 
the surrounding gas and produce outflows. 
The sub-grid multiphase model implemented in the SPH runs considers two different phases for the gas particles named cold and hot. 
According to the SNe energy that a cold gas particle receives and accumulates, it can be \textit{promoted} to 
a hot gas phase particle where the energy received is converted into internal energy that at the same time, can produce mass-loaded
galactic winds \citep[see,][for further details]{Scannapieco2005,Scannapieco2006}.
Therefore, the promoted gas particles can give us information about the galactic outflows and at the same time, 
the star-forming evolution of the galaxy.
In Fig.~\ref{fig:promoted} we show the temperature of the promoted gas particles dumped to the hot phase, for the central galaxies 
within the SPH haloes, as a function of redshift.
We present our results from the four haloes for the full sample of promoted particles (dotted lines), and for the promoted gas particles with temperatures above T$>10^6$~K (filled lines). 
In agreement with the results from \citet{Scannapieco2009}, we find that the promoted gas particles from the galaxies in Aq-A-SPH
and Aq-C-SPH have a main burst between $z \sim 3.0 - 3.5$, as consequence of the prior star formation in these haloes.
The haloes Aq-D-SPH and Aq-E-SPH show an increase of the mean temperature at lower redshifts, around $z\sim 1.5$ which is not significantly important compared with
the star-forming bursts in Aq-A-SPH and Aq-C-SPH.
These findings are in agreement with the different stellar mass evolution found in Fig.\ref{fig:MAH-bar} and Fig.~\ref{fig:cumularive-MD}. 

To investigate the contraction of the dark matter haloes due to galaxy assembly, we implement the fiducial adiabatic contraction (AC) model
developed by \citet{Blumenthal1986}, the alternative proposed by \citet{Gnedin2004} and two empirical models presented in \citet{Abadi2010},
and \citet{Pedrosa2010}.
The adiabatic contraction (AC) model is based on the fact that the condensation of baryons during galaxy assembly affects the mass profile of the dark matter halo.
The AC assume that the system changes its gravitational potential slowly which in turn can be approximated as adiabatic. In the fiducial model, the gravitational potential
is symmetric, the orbits are circular and the angular momentum of the particles is conserved. Hence, the fiducial AC from \citet{Blumenthal1986} assumes
that the quantity $M(r) r$ is conserved.
\citet{Gnedin2004} propose a modified AC model based on conservation of the quantity $M(\overline{r}) r$, where
$M(\overline{r})$ is the mass enclosed within the orbit-averaged radius. 
Other alternative (empirical) models are those proposed by \citet{Abadi2010} and \citet{Pedrosa2010}, following empirical formulas to reproduce the 
results from different cosmological simulations at $z=0$.
In Fig.~\ref{fig:AC} we implement the aforementioned AC models to the Aquarius haloes at redshifts $z=0,$ 1, 2, 3 and 4, by computing the dark matter mass 
in the inner region, within 5 percent the virial radius at each redshift (5\%R$_{200,z}$).
We find that between $z=2-4$, the haloes Aq-A-SPH and Aq-C-SPH show a smaller contraction than Aq-D-SPH and Aq-E-SPH. This 
effect might be due to the outflows produced in Aq-A-SPH and Aq-C-SPH as we show in Fig.~\ref{fig:promoted}.

In Fig.~\ref{fig:AC} we find that the model
proposed by \citet{Gnedin2004} is the one that best match the dark matter mass in the inner regions of the SPH haloes in all the redshift
range studied. The fiducial model from \citet{Blumenthal1986} shows an excess of the dark matter mass in the inner regions, suggesting that 
this model overpredicts the effect of the baryons.
The models proposed by \citet{Abadi2010} and \citet{Pedrosa2010} have a good agreement at $z=0$ but deviates 
and  underpredicts the dark matter mass within the inner regions as redshift increases. 
Hence, our findings suggest that the model from \citet{Gnedin2004} is the one that best reproduces the impact
of baryons in the inner regions of the SPH Aquarius haloes between $z=0-4$. 

Recently, \citet{Dutton2016} investigate a sample of 100 haloes
at $z=0$ between M$_{200} = 10^{10}$-$10^{12}$~M$_{\odot}$, finding that the halo response is correlated with 
the star formation efficiency and the compactness of the stellar system \citep[see also][]{Pedrosa2009,Pedrosa2010}.
In particular, their results suggest that galaxies with high star formation efficiencies 
($\epsilon_{\rm SF} = (M_*/M_{200})/(\Omega_{\rm b}/\Omega_{\rm m}$) $> 0.5$), which is the case of our SPH sample,
follow a contraction that is well described by \citet{Gnedin2004}. This result is in agreement with our findings.

The halo Aq-C-M14 shows a better agreement with the empirical models of \citet{Abadi2010} and \citet{Pedrosa2010}
for all the redshift range studied.
We also compare the mass distribution of the four AC models implemented, between 3\%R$_{200,z}$--R$_{200,z}$
finding similar results than those obtained for the inner regions (see Appendix~\ref{sec:appx}).

\section{Summary and conclusions}\label{sec:conclusions}

We have investigated the impact of baryons onto four Milky Way-type dark matter haloes from the Aquarius project, between $z=0-4$.
In order to disentangle the effect of baryons, we compare the SPH runs of \citet{Scannapieco2009} with the corresponding
DMO runs.
The SPH run includes star formation, a multiphase model for the gas component with metal-dependent cooling, 
and phase-dependent  treatments of SNe feedback and chemical enrichment. 
We note here, that some of the SPH haloes have a higher stellar mass to halo mass content than the values typically reported by 
 abundance matching techniques (see Fig.~\ref{fig:SMHM_Comparison}). Hence the galaxies within these haloes are likely representing
a particular subset of real galaxies.
Hence, in order to check the effect of sub-grid physics and numerical artefacts,
we also analyse a version of Aq-C run by \citep{Marinacci2014} using the {\sc arepo} code. This version properly reproduce the results
reported by abundance matching techniques at $z=0$.

As expected, we find that the evolution of the dark matter haloes is affected by the presence of baryons.
For the SPH haloes, we fit the dark matter halo profiles at $z=0$ with the model proposed 
by \citet{DiCintio2014} which accounts for the impact of baryons 
considering that three of the parameters of the functional form depend on
the star formation efficiency. 
Since the star formation efficiencies of the SPH haloes are higher than the 
ranges analysed by \citet{DiCintio2014}, we relax the constrains that the model has on the concentration 
and the halo mass. We propose two methods to implement this profile, first by fitting the concentration, and second by fitting the concentration 
and the halo mass. 
Our results show that the model proposed by \citet{DiCintio2014} present good agreement in 
the inner parts, but overpredicts the density in the outskirts of some Aquarius haloes (see Fig.~\ref{fig:fits-profile}).
This might be due to the impact of recent mergers in the outer regions of the haloes. 

We also fit the SPH haloes at $z=0, 1, 2, 3$ and 4 using the Einasto profile.
Our results show that all the SPH haloes are slightly better fitted by the Einasto model than \citet{DiCintio2014}.
However it is important to bear in mind that Einasto contains an extra free parameter.
Regardless, both the DC14 and the Einasto models provide a very good agreement on recovering
the correct halo mass, with 1-10 percent differences between the true mass and the fit mass.

We find differences in the evolution on the dark matter halo profiles for the SPH haloes.
Two of the haloes present weak evolution in the range studied (Aq-A-SPH and Aq-C-SPH), while
the other two show a significant evolution with redshift between $z\sim 1-4$ (Aq-D-SPH and Aq-E-SPH, see Fig.~\ref{fig:DM-profiles2}).
The halo run with a different code and subgrid scheme (Aq-C-M14) shows
similar evolution than Aq-C-SPH.

The mass accretion history (MAH) of the dark matter haloes shows that SPH haloes assemble earlier than their DMO counterparts.
This can be explained since baryons make the haloes more concentrated and in turn more massive.
From the MAH of dark matter and baryons in the SPH haloes, we find that there is a connection between the mass assembly 
and the evolution of the dark matter profiles.
Hence, the SPH haloes reach stable dark matter profiles in time when they attain 
the 80\% of their present baryonic mass. The SPH haloes also show that 
$\sim 60$\% of the baryonic mass at present is composed by stars at this stage (see Fig.~\ref{fig:MAH}~and~\ref{fig:MAH-bar}).
This amount of baryonic mass is achieved earlier for the SPH haloes that show slight evolution of the
dark matter density profiles.
This is also evidenced in the cumulative stellar mass distribution which is different 
between these two set of haloes.
Hence, our findings suggest that the stellar mass content 
is a key ingredient that shapes and makes the internal region of the haloes more stable.
Furthermore, our results confirm that the stellar mass content depends on the halo assembly history. 
For the halo Aq-C-M14 we also find a close connection between the MAH, the amount of baryons, and the evolution of the dark
matter density profiles.
This halo reaches a stable dark matter profile at the time it attains
the 80\% of the present baryonic mass.
This is in agreement with the findings for the SPH runs although
the time at which this fraction is attained vary with the systems,
for different assembly histories and different subgrid physics. 
The main difference with respect to the SPH haloes is that Aq-C-M14 is composed mainly by gas at high redshift, 
reaching 60\% gas mass of the present baryonic mass at $z\sim 1.5$.
This is the expected behaviour because of the more efficient feedback which contribute to the regulation 
of the star formation activity, delaying it towards lower redshifts.
Nonetheless at high redshift most of the current models overpredicts
the amount of stellar mass within the haloes. This aspect should be tackled in the future \citep[see]{Aumer2013,Artale2015,Grand2017}.
 
The global shape of the dark matter haloes show that both DMO and SPH counterparts
are more prolate at high redshift and turn to triaxial/oblate at low redshift.
The inner shape of the haloes present a clear different behaviour, finding that SPH haloes
are more oblate than their DMO counterparts. This trend is also present in Aq-C-M14.
This result is expected due to the impact of baryons in the inner region (see Fig.~\ref{fig:TriaxRvir}). 

We also investigate the specific angular momentum in the inner region of these haloes 
within five percent the virial radius through Lagrangian and Eulerian selection.
In particular, for the Lagrangian-selection, the DMO haloes loss $\sim 85 - 95$\% of its angular momentum between $z=4$ to $z=0$.
This result indicates that dark matter halo particles loss angular momentum and concentrate
in the central region of the haloes, in agreement with our findings for the halo profiles.
The SPH haloes that reach stable dark matter profiles earlier show that the specific angular momentum is conserved in time
in the Lagrangian-selection. 
The haloes that show a contraction of the dark matter profile
present a loss of the specific angular momentum in time. However, this loss is lower than what we
find for their DMO counterparts.
The specific angular momentum for baryons from Lagrangian-selection show similar trends than those
obtained from the dark matter particles, showing that the specific angular momentum is 
roughly conserved when the SPH haloes present stable dark matter density profiles.
The halo Aq-C-M14 show also that the specific angular momentum in the inner region is roughly constant, with a slight
drop at low redshift. 

To quantify the impact of baryons onto the contraction of the dark matter haloes, we implement four adiabatic contraction models. 
The fiducial one from \citet{Blumenthal1986}
and the empirical models from \citet{Gnedin2004}, \citet{Pedrosa2010} and \citet{Abadi2010}. 
Our findings show that the model proposed by \citet{Gnedin2004} is the best representation of the inner mass (5\%R$_{200,z}$)
content of the SPH haloes between $z=0-4$.
However, this is not the case for Aq-C-M14 which is in better agreement with \citet{Pedrosa2010} and \citet{Abadi2010}
in all the redshift range studied. This is due to the different evolution of the stellar mass content in the 
Aq-C-M14 compared with the SPH haloes, since the star formation activity is more strongly regulated and delayed to lower redshifts.

We have shown that the baryonic mass content within the central region of the dark matter haloes 
is a key ingredient in determining the stability and concentration of the haloes.
We find that how these baryons are put together is also an important aspect.
In the future, a statistical sample should be used to quantify in detail these differences.

\section*{Acknowledgement}
We would like to thank Aaron Ludlow and Julio Navarro for providing access to the dark matter only runs of the Aquarius project, 
and to Facundo Gomez, Volker Springel and Federico Marinacci for providing the snapshots of the run of Aq-C-4 corresponding to
\citet{Marinacci2014}. We would also like to thank Cecilia Scannapieco for useful discussions.
The project was partially supported by Fondecyt Regular 1153305 and Nucleo Milky Way Project of UNAB, and by
PICT 2011-0959 and PIP 2012-0396 (Mincyt, Argentina).
The authors acknowledge the Southern Astrophysics Network Proyecto Redes 150078 (Conicyt-Chile) for
partial financial support.
MCA acknowledges financial support from the Austrian National Science Foundation through FWF stand-alone grant P31154-N27.
ADC acknowledges financial support from  a Marie-Sk\l{}odowska-Curie Individual Fellowship grant, H2020-MSCA-IF-2016 Grant agreement 748213 DIGESTIVO.

\bibliographystyle{aa} 

\bibliography{Aquarius}

\appendix
\section{Adiabatic contraction}
\label{sec:appx}

\begin{figure*}
  \centering
  \includegraphics[width=0.4\textwidth]{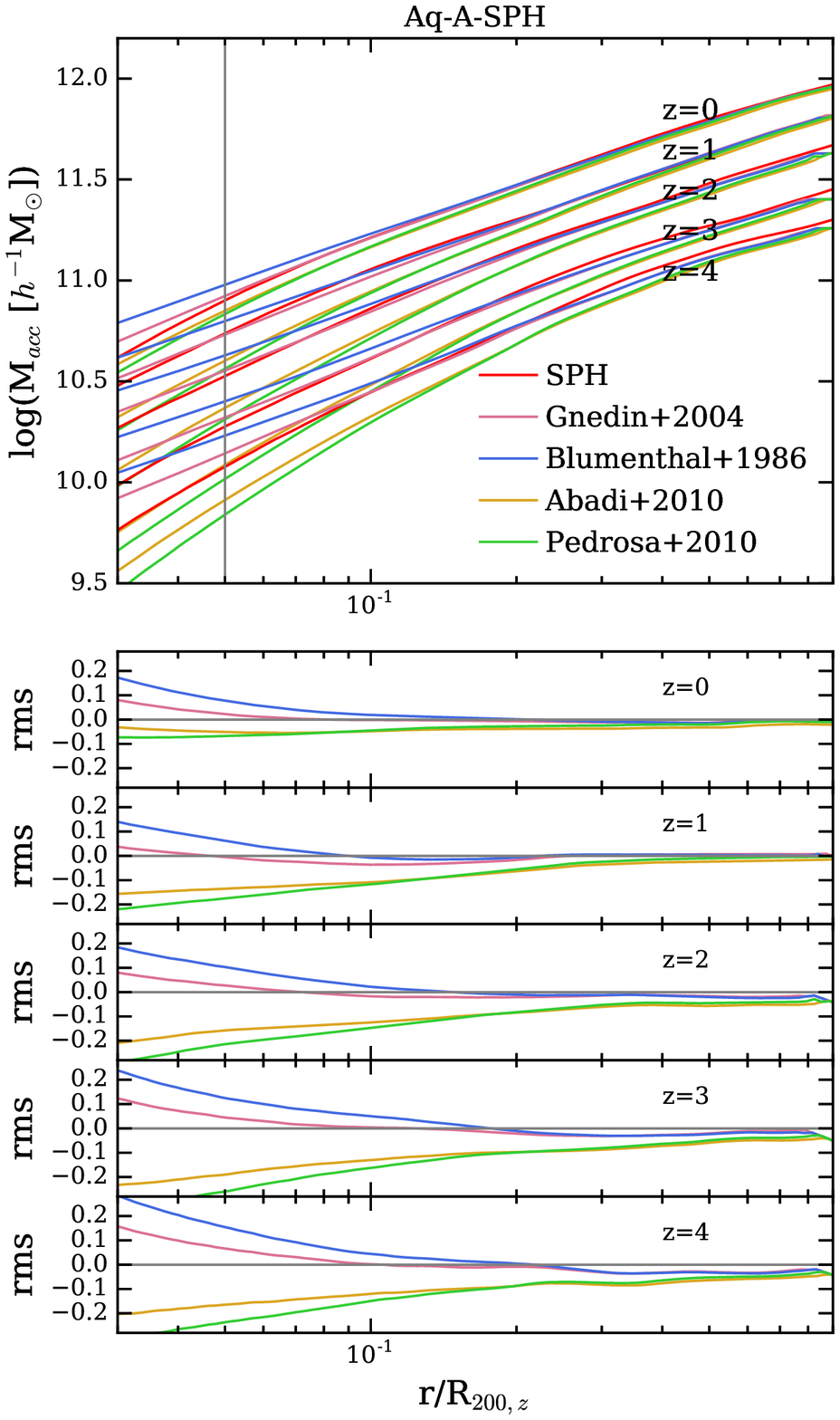}
  \includegraphics[width=0.4\textwidth]{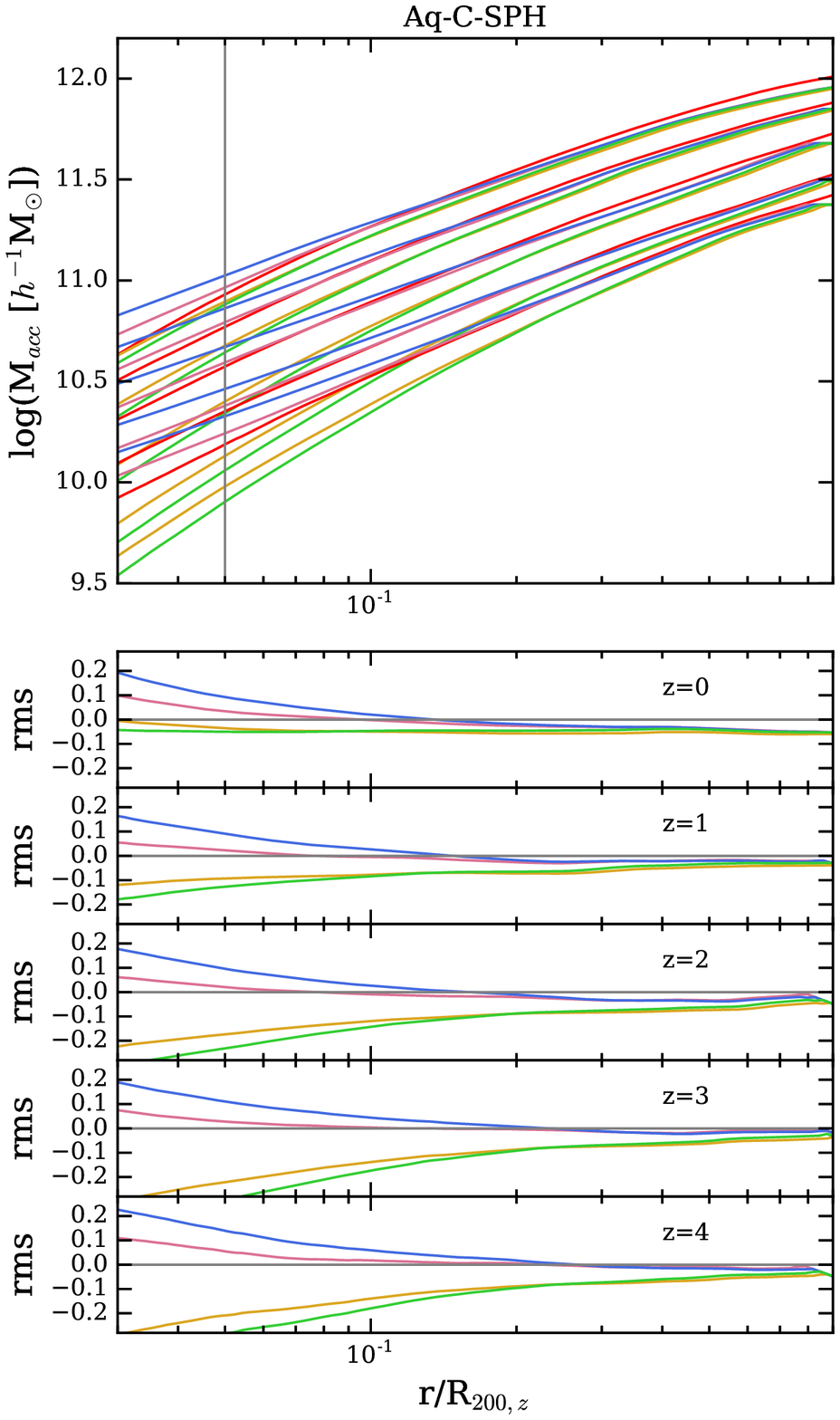}
  \includegraphics[width=0.4\textwidth]{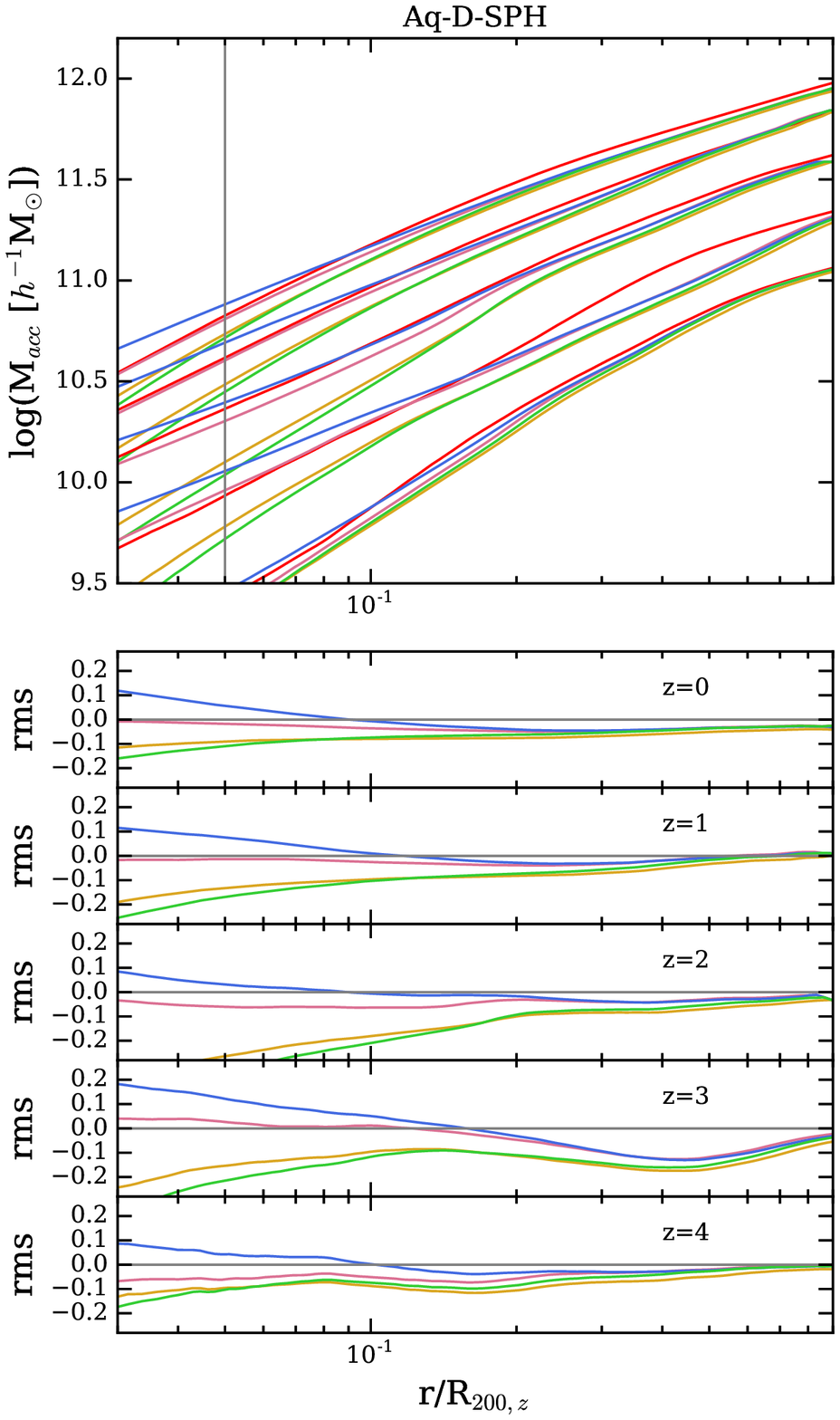}
  \includegraphics[width=0.4\textwidth]{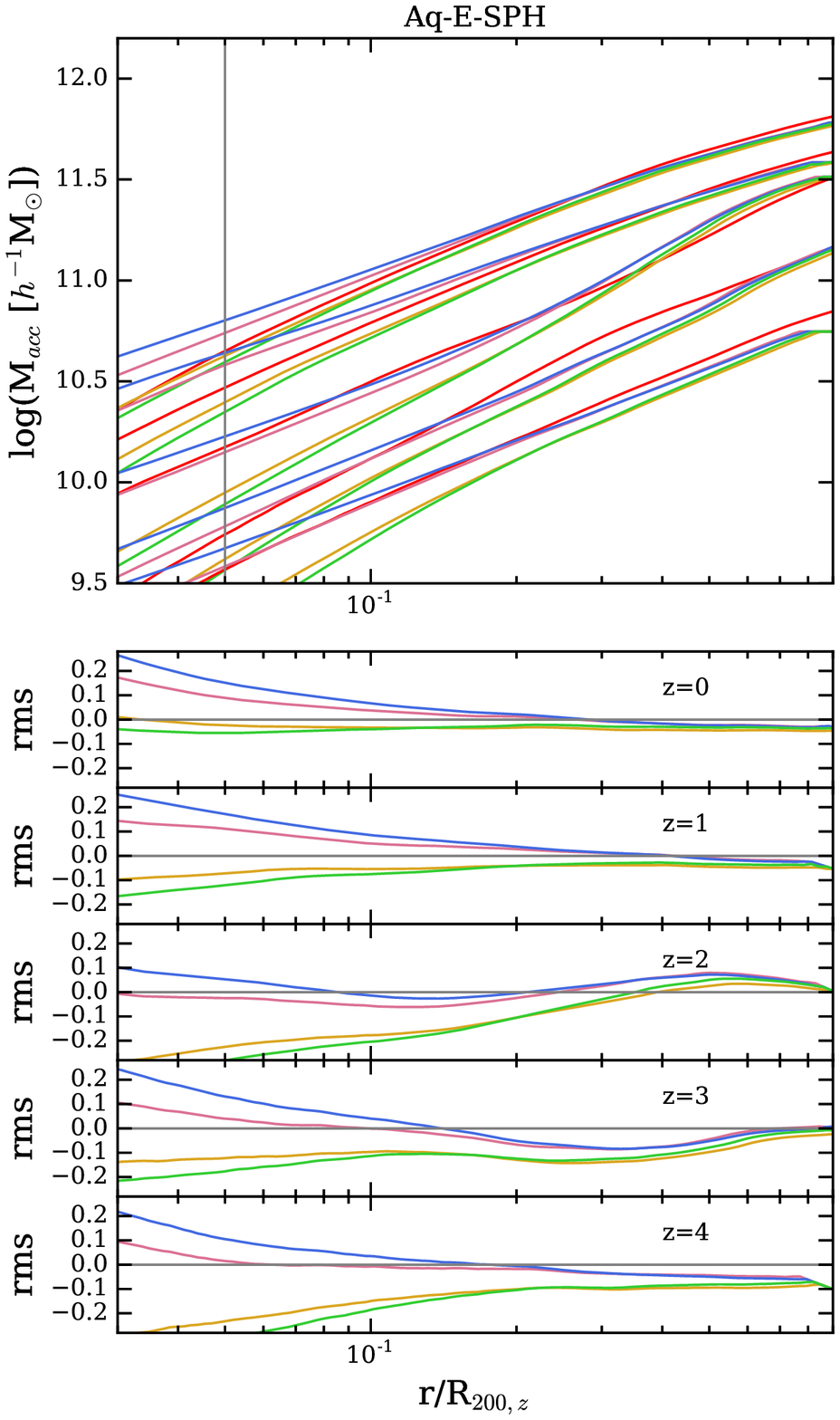}
 \caption{Mass distribution obtained from the four SPH Aquarius haloes,
 compared with four different halo contraction implementations
 at $z=0, 1, 2, 3,$ and 4 (from top to bottom for each halo). The AC models are 
  \citet{Blumenthal1986} (blue lines), \citet{Gnedin2004} (pink lines). We also compare with
  the empirical models from \citet{Abadi2010} (orange lines)
  and \citet{Pedrosa2010} (green lines). 
  Vertical grey line indicates the 5\% virial radius at each redshift.
  We also include the residuals between the mass distribution of the SPH haloes
  with each AC model.}
  \label{fig:Apx_AC}
 \end{figure*}

In Fig.~\ref{fig:Apx_AC}, we present the mass distribution of the SPH haloes between 
3\%R$_{200,z}$--R$_{200,z}$ compared with four halo contraction models  
(\citet{Blumenthal1986} blue lines, \citet{Gnedin2004} pink lines, and empirical models of \citet{Abadi2010} orange lines,
and \citet{Pedrosa2010} green lines) at $z=0$, 1, 2, 3, and 4.
The distances are normalized by the virial radius at each redshift, R$_{200,z}$.
We also include the residuals between the models and the mass distribution obtained from the SPH haloes.
Vertical grey indicates the 5\%R$_{200,z}$ which is the distance adopted in this
work to explore the inner regions of the halo.
As we state in Sec.~\ref{sec:ac}, at $z=0$ the models proposed by \citet{Abadi2010} and \citet{Pedrosa2010}
are in good agreement with the mass distribution of the SPH haloes. However, the model from \citet{Gnedin2004}
is the best reproducing the mass distribution in all the redshift range studied.

\IfFileExists{\jobname.bbl}{}
{\typeout{}
\typeout{****************************************************}
\typeout{****************************************************}
\typeout{** Please run "bibtex \jobname" to optain}
\typeout{** the bibliography and then re-run LaTeX}
\typeout{** twice to fix the references!}
\typeout{****************************************************}
\typeout{****************************************************}
\typeout{}
}
\end{document}